\documentclass{article}
\usepackage{arxiv}
\usepackage[T1]{fontenc}
\usepackage[latin9]{inputenc}
\usepackage{geometry}

\usepackage{latexsym}
\usepackage{graphicx}
\usepackage{amsfonts,amssymb,amsmath}
\usepackage{hyperref}

\usepackage{color}
\usepackage{subcaption}
\usepackage{hyperref}

\usepackage{tikz}
\usepackage{pgfplots}
\pgfplotsset{compat=1.16}
\usetikzlibrary{spy,calc}

\newif\ifblackandwhitecycle
\gdef\patternnumber{0}
\pgfkeys{/tikz/.cd,
    zoombox paths/.style={
        draw=orange,
        very thick
    },
    black and white/.is choice,
    black and white/.default=static,
    black and white/static/.style={ 
        draw=white,   
        zoombox paths/.append style={
            draw=white,
            postaction={
                draw=black,
                loosely dashed
            }
        }
    },
    black and white/static/.code={
        \gdef\patternnumber{1}
    },
    black and white/cycle/.code={
        \blackandwhitecycletrue
        \gdef\patternnumber{1}
    },
    black and white pattern/.is choice,
    black and white pattern/0/.style={},
    black and white pattern/1/.style={    
            draw=white,
            postaction={
                draw=black,
                dash pattern=on 2pt off 2pt
            }
    },
    black and white pattern/2/.style={    
            draw=white,
            postaction={
                draw=black,
                dash pattern=on 4pt off 4pt
            }
    },
    black and white pattern/3/.style={    
            draw=white,
            postaction={
                draw=black,
                dash pattern=on 4pt off 4pt on 1pt off 4pt
            }
    },
    black and white pattern/4/.style={    
            draw=white,
            postaction={
                draw=black,
                dash pattern=on 4pt off 2pt on 2 pt off 2pt on 2 pt off 2pt
            }
    },
    zoomboxarray inner gap/.initial=5pt,
    zoomboxarray columns/.initial=2,
    zoomboxarray rows/.initial=2,
    subfigurename/.initial={},
    figurename/.initial={zoombox},
    zoomboxarray/.style={
        execute at begin picture={
            \begin{scope}[
                spy using outlines={%
                    zoombox paths,
                    width=\imagewidth / \pgfkeysvalueof{/tikz/zoomboxarray columns} - (\pgfkeysvalueof{/tikz/zoomboxarray columns} - 1) / \pgfkeysvalueof{/tikz/zoomboxarray columns} * \pgfkeysvalueof{/tikz/zoomboxarray inner gap} -\pgflinewidth,
                    height=\imageheight / \pgfkeysvalueof{/tikz/zoomboxarray rows} - (\pgfkeysvalueof{/tikz/zoomboxarray rows} - 1) / \pgfkeysvalueof{/tikz/zoomboxarray rows} * \pgfkeysvalueof{/tikz/zoomboxarray inner gap}-\pgflinewidth,
                    magnification=3,
                    every spy on node/.style={
                        zoombox paths
                    },
                    every spy in node/.style={
                        zoombox paths
                    }
                }
            ]
        },
        execute at end picture={
            \end{scope}
            \node at (image.north) [anchor=north,inner sep=0pt] {\subcaptionbox{\label{\pgfkeysvalueof{/tikz/figurename}-image}}{\phantomimage}};
            \node at (zoomboxes container.north) [anchor=north,inner sep=0pt] {\subcaptionbox{\label{\pgfkeysvalueof{/tikz/figurename}-zoom}}{\phantomimage}};
     \gdef\patternnumber{0}
        },
        spymargin/.initial=0.5em,
        zoomboxes xshift/.initial=1,
        zoomboxes right/.code=\pgfkeys{/tikz/zoomboxes xshift=1},
        zoomboxes left/.code=\pgfkeys{/tikz/zoomboxes xshift=-1},
        zoomboxes yshift/.initial=0,
        zoomboxes above/.code={
            \pgfkeys{/tikz/zoomboxes yshift=1},
            \pgfkeys{/tikz/zoomboxes xshift=0}
        },
        zoomboxes below/.code={
            \pgfkeys{/tikz/zoomboxes yshift=-1},
            \pgfkeys{/tikz/zoomboxes xshift=0}
        },
        caption margin/.initial=4ex,
    },
    adjust caption spacing/.code={},
    image container/.style={
        inner sep=0pt,
        at=(image.north),
        anchor=north,
        adjust caption spacing
    },
    zoomboxes container/.style={
        inner sep=0pt,
        at=(image.north),
        anchor=north,
        name=zoomboxes container,
        xshift=\pgfkeysvalueof{/tikz/zoomboxes xshift}*(\imagewidth+\pgfkeysvalueof{/tikz/spymargin}),
        yshift=\pgfkeysvalueof{/tikz/zoomboxes yshift}*(\imageheight+\pgfkeysvalueof{/tikz/spymargin}+\pgfkeysvalueof{/tikz/caption margin}),
        adjust caption spacing
    },
    calculate dimensions/.code={
        \pgfpointdiff{\pgfpointanchor{image}{south west} }{\pgfpointanchor{image}{north east} }
        \pgfgetlastxy{\imagewidth}{\imageheight}
        \global\let\imagewidth=\imagewidth
        \global\let\imageheight=\imageheight
        \gdef\columncount{1}
        \gdef\rowcount{1}
        
    },
    image node/.style={
        inner sep=0pt,
        name=image,
        anchor=south west,
        append after command={
            [calculate dimensions]
            node [image container,subfigurename=\pgfkeysvalueof{/tikz/figurename}-image] {\phantomimage}
            node [zoomboxes container,subfigurename=\pgfkeysvalueof{/tikz/figurename}-zoom] {\phantomimage}
        }
    },
    color code/.style={
        zoombox paths/.append style={draw=#1}
    },
    connect zoomboxes/.style={
    spy connection path={\draw[draw=none,zoombox paths] (tikzspyonnode) -- (tikzspyinnode);}
    },
    help grid code/.code={
        \begin{scope}[
                x={(image.south east)},
                y={(image.north west)},
                font=\footnotesize,
                help lines,
                overlay
            ]
            \foreach \x in {0,1,...,9} { 
                \draw(\x/10,0) -- (\x/10,1);
                \node [anchor=north] at (\x/10,0) {0.\x};
            }
            \foreach \y in {0,1,...,9} {
                \draw(0,\y/10) -- (1,\y/10);                        \node [anchor=east] at (0,\y/10) {0.\y};
            }
        \end{scope}    
    },
    help grid/.style={
        append after command={
            [help grid code]
        }
    },
}

\newcommand\phantomimage{%
    \phantom{%
        \rule{\imagewidth}{\imageheight}%
    }%
}
\newcommand\zoombox[2][]{
    \begin{scope}[zoombox paths]
        \pgfmathsetmacro\xpos{
            (\columncount-1)*(\imagewidth / \pgfkeysvalueof{/tikz/zoomboxarray columns} + \pgfkeysvalueof{/tikz/zoomboxarray inner gap} / \pgfkeysvalueof{/tikz/zoomboxarray columns} ) + \pgflinewidth
        }
        \pgfmathsetmacro\ypos{
            (\rowcount-1)*( \imageheight / \pgfkeysvalueof{/tikz/zoomboxarray rows} + \pgfkeysvalueof{/tikz/zoomboxarray inner gap} / \pgfkeysvalueof{/tikz/zoomboxarray rows} ) + 0.5*\pgflinewidth
        }
        \edef\dospy{\noexpand\spy [
            #1,
            zoombox paths/.append style={
                black and white pattern=\patternnumber
            },
            every spy on node/.append style={#1},
            x=\imagewidth,
            y=\imageheight
        ] on (#2) in node [anchor=north west] at ($(zoomboxes container.north west)+(\xpos pt,-\ypos pt)$);}
        \dospy
        \pgfmathtruncatemacro\pgfmathresult{ifthenelse(\columncount==\pgfkeysvalueof{/tikz/zoomboxarray columns},\rowcount+1,\rowcount)}
        \global\let\rowcount=\pgfmathresult
        \pgfmathtruncatemacro\pgfmathresult{ifthenelse(\columncount==\pgfkeysvalueof{/tikz/zoomboxarray columns},1,\columncount+1)}
        \global\let\columncount=\pgfmathresult
        \ifblackandwhitecycle
            \pgfmathtruncatemacro{\newpatternnumber}{\patternnumber+1}
            \global\edef\patternnumber{\newpatternnumber}
        \fi
    \end{scope}
}

\usepackage{babel}
\usepackage{cite}
\usepackage{booktabs} 
\usepackage{xcolor} 
\usepackage{fancyvrb} 

\newcommand\figref{Fig.~\ref}

\title{A guided edge-aware smoothing-sharpening filter based on patch interpolation
model and generalized Gamma distribution}

\author{
  DaGuang~Deng \\
  Department of  Engineering \\
  La Trobe University\\
  Bundoora, VIC 3086, Australia\\
  \texttt{d.deng@latrobe.edu.au} \\
   \And
 Fernando~J.~Galetto\thanks{Corresponding author.}\\
  Department of  Engineering \\
  La Trobe University\\
  Bundoora, VIC 3086, Australia\\
  \texttt{f.galetto@latrobe.edu.au} \\
     \And
 Mukhalad~Al--nasrawi\\
  Electrical Power Engineering \\
  Al-Furat Al-Awsat Technical University\\
  Al-Mussaib, Iraq.\\
  \texttt{com.muk@atu.edu.iq} \\
     \And
 Waseem~Waheed\\
  Department of  Engineering \\
  La Trobe University\\
  Bundoora, VIC 3086, Australia\\
  \texttt{w.waheed@latrobe.edu.au} \\
}

\begin{document}
\maketitle

\begin{abstract}
 Smoothing and sharpening are two fundamental image processing operations. The latter is usually related to the former through the unsharp masking algorithm. In this paper, we develop a new type of filter which performs smoothing or sharpening via a tuning parameter. The development of the new filter is based on (1) a new Laplacian-based filter formulation which unifies the smoothing and sharpening operations, (2) a patch interpolation model similar to that used in the guided filter which provides edge-awareness capability, and (3) the generalized Gamma distribution which is used as the prior for parameter estimation. We have conducted detailed studies on the properties of two versions of the proposed filter (self-guidance and external guidance). We have also conducted experiments to demonstrate applications of the proposed filter. In the self-guidance case, we have developed adaptive smoothing and sharpening algorithms based on texture, depth and blurriness information extracted from an image. Applications include enhancing human face images, producing shallow depth of field effects, focus-based image enhancement, and seam carving. In the external guidance case, we have developed new algorithms for combining flash and no-flash images and for enhancing multi-spectral images using a panchromatic image.
\end{abstract}


\section{Introduction}

Smoothing and sharpening are two fundamental operations in image processing. Traditionally, smoothing is used to reduce noise, while sharpening is used to enhance details \cite{gonzalez2018}. In recent years, smoothing has found increasingly more applications in graphics, computational photography, and computer vision. Edge-aware smoothing,  which preserves sharp edges of objects, has been actively studied. Well known edge-aware filters include  \cite{Tomasi1998, he2012guided, deng2016guided, wei2018joint, yin2019side, Sun2020}. Among them, the guided filter \cite{he2012guided} and its weighted versions \cite{li2015weighted, gradientDomainGF, Sun2020} have the advantage of a low computational complexity of $\mathcal{O}(N)$ in addition to their good performance. Applications of edge-aware filters include detail enhancement, flash no-flash image denoising, upsampling of depth map, image abstraction, image dehazing, tone mapping and contrast enhancement to name a few. These applications have been studied extensively in the literature. 

On the other hand, sharpening is usually achieved through the unsharp masking algorithm \cite{gonzalez2018}. Let $I$ and $S$ be the observed image to be processed and the sharpened image, respectively. The sharpened image is produced by $S=I+\gamma Z$ where $\gamma$ is called the sharpening gain and $Z$ is the output of a high-pass filter. The use of a linear filter to produce $Z$ presents two main disadvantages in some applications: high sensitivity to noise and halo artifacts due to overshoot in high contrast regions. Non-linear filters were proposed in \cite{mitra1991new, ramponi1998cubic, ramponi1996nonlinear} to reduce the effect of noise but not solving the halo artifacts. Edge-aware filters are the main tools to combat the halo effect. In addition, there are many studies on using adaptive gain to perform content adaptive sharpening. For example, in \cite{polesel2000image}, a pixel adaptive gain $\gamma$ based on the dynamics of the image is proposed to sharpen areas of mid-range contrast, to avoid overshooting in high contrast regions, and to produce minimal sharpening at smooth regions. Attempts have also been made to formulate the adaptive sharpening problem as an optimization problem \cite{ContentAdaptiveImageDetailEnhancement}. A similar approach was taken by \cite{ye2018blurriness} which used the local blurriness to vary the sharpening gain $\gamma$. The main goal is to avoid sharpening the very smooth background which is intentionally produced by the photographer to achieve the effect of shallow depth of field.


Although smoothing and sharpening are related through $Z=I-J$, where $J$ is a smoothed version of $I$, and edge-aware filters are used to produce $J$ to minimize the halo effect, these two operations are usually used in different applications. The main motivation of this work is to develop a unified framework such that smoothing and sharpening can be integrated in one filter whose function can be controlled by varying a parameter. Our aim is to develop a filter with a tuning parameter such that when it is set smaller/greater than 1 the filter is smoothing/sharpening. The filter must also have the edge-awareness property such that it does not blur edges when used in smoothing mode and does not create halos when used in sharpening mode. A distinctive advantage of such formulation is that it allows the user to perform selective smoothing and sharpening in different areas of an image to produce results such as smoothing the background while sharpening the main object. In addition, the unification of these two operations in one filter allows the user to have a better control in information fusion applications such as Pan-sharpening \cite{shah2008efficient} and flash-no-flash imaging \cite{he2012guided}.

The main contributions of this work and organization of this paper are summarized in the following.
\begin{itemize}

\item A systematic formulation of a new type of filter (section \ref{subsec:The-smoothing-sharpening-filter}) based on the Laplacian operator, which unifies smoothing and sharpening operations in one filter. The function and level of smoothing or sharpening are controlled by varying the value of a parameter.

\item The development of an edge-aware smoothing-sharpening filter (section \ref{sec:The-proposed-smoothing-sharpenin}) based on a patch interpolation model similar to that of the guided filter. Parameters of the filter are determined by minimizing the negative posterior probability. The generalized Gamma distribution is used as the prior. The filter includes the original guided filter as a special case. Self-guidance and external-guidance versions of the filter have been developed and their properties are analyzed. Both versions of the filter are of the same computational complexity as that of the guided filter.

\item We have demonstrated the performance of the proposed filter in a number of applications in section \ref{App}. Using the self-guided filter, we have developed adaptive smoothing-sharpening algorithms by extracting information of texture, depth, and blurriness to adjust the filter parameter to achieve content-aware processing. Applications include enhancement of images of human face, creating the effect of shallow depth of field, smoothing/sharpening guided by blurriness, and pre-processing an image to achieve better seam carving results. Using the filter in external guidance, we have applied the filter to combine images taken under flash and no-flash conditions, producing much better results than those produced by the guided filter. We have applied the filter to solve the Pan-sharpening \cite{Qu2017} problem which combines information from multi-spectral images with a panchromatic image. Both subjective and objective comparison are discussed to validate the applications of the proposed filter.
\end{itemize}

\section{The smoothing-sharpening filter, the self-guided filter, and main
ideas of this work}

In this section, we first present a filter called the smoothing-sharpening
filter in which smoothing and sharpening is configured through the
setting of a parameter. Next, we revisit the basic idea of the guided
filter. We then discuss the main idea of the proposed edge-aware smoothing-sharpening
filter.

\subsection{The smoothing-sharpening filter\label{subsec:The-smoothing-sharpening-filter}}

We develop a unified framework for combining 
smoothing and sharpening into one filter. We first define the filter
as
\begin{equation}
    J(n)=I(n)+(1-\alpha)\Delta I(n)\label{eq:smoothing}
\end{equation}
where $I(n)$ and $J(n)$ are pixels of the input and output images at location $n$, $\alpha$ is a parameter, and $\Delta$ is the discrete Laplace operator (which is referred to as the Laplacian operator in rest of this paper) defined as
\begin{equation}
    \Delta I(n)=\mu(n)-I(n)\label{eq:laplacian}
\end{equation}
where $\mu(n)=\frac{1}{N}\sum_{m\in\Omega_{n}}I(m)$ is the mean of
the image calculated over a patch centered at location $n$. The
symbol $\Omega_{n}$ represents the set of pixel indices of the patch
and $N=|\Omega_{n}|$ is the number of pixels. The particular parameterization of this filter by $1-\alpha$ will become clear in the following discussions.

\begin{figure}
    \begin{centering}
    \includegraphics[scale=0.65]{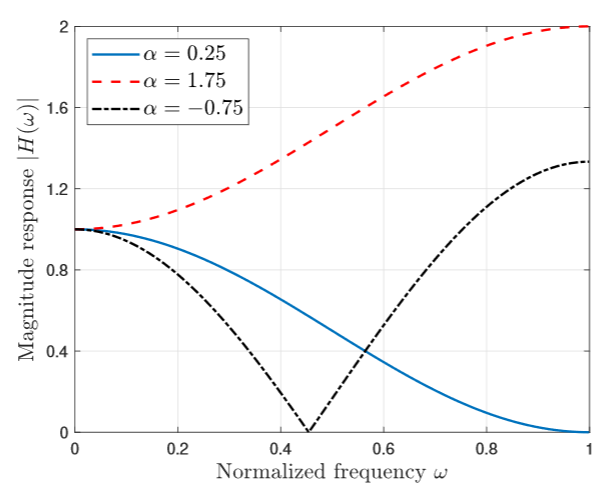}
    \par\end{centering}
    \caption{\label{fig:Magnitude-response-of}Magnitude response of the smoothing-sharpening
    filter. When $0<\alpha<1$, it is a low-pass filter. When $\alpha>1$,
    it is a high-frequency emphasis filter. When $\alpha<0$, the filter's function is not well defined.}
\end{figure}

To demonstrate the characteristics of this filter, we consider a simple example of
a 1-D filter of which the mean is calculated by an average filter  $\mu(n)=\frac{1}{3}\sum_{m=-1}^{1} I(n-m)$. The
impulse response of the filter stated in \eqref{eq:smoothing} is
then given by $h(n)=\{(1-\alpha),(1+2\alpha),(1-\alpha)\}/3$. The
frequency response is calculated as follows
\begin{equation}
    H(\omega)=\sum_{n=-1}^{1}h(n)e^{-j\omega n}=1-\frac{4(1-\alpha)}{3}\sin^{2}\frac{\omega}{2}
\end{equation}
In \figref{fig:Magnitude-response-of}, we plot the magnitude response
($|H(\omega)|$) of this filter for three settings of $\alpha$.
We can see that (a) when $0<\alpha<1$, it is a low-pass filter,
(b) when $\alpha>1$, it is a high-frequency emphasis filter, and
(c) when $\alpha<0$, the filter's function is not well defined and is not considered in this paper. 

Substitution of \eqref{eq:laplacian} into \eqref{eq:smoothing}, we have 
\begin{equation}
    J(n)=\alpha I(n)+(1-\alpha)\mu(n)\label{eq:the filter}
\end{equation}
which is a weighted average between $I(n)$ and $\mu(n)$ when $0\le\alpha<1$. It is a low-pass filter.
On the other hand, when $\alpha > 1$ we can
then re-write \eqref{eq:the filter} as
\begin{equation}
    J(n) =\mu(n)+\alpha (I(n)-\mu(n))\label{eq:s1}\\
\end{equation}
Equation \eqref{eq:s1} is the unsharp masking operation which is
a sharpening filter. The equivalence of \eqref{eq:s1} and \eqref{eq:smoothing}
shows that the filter stated in \eqref{eq:smoothing} can be configured
as either a smoothing filter $0<\alpha<1$ or a sharpening filter $\alpha>1$.
We can further re-write the filter stated in \eqref{eq:the filter}
in the following equivalent form
\begin{equation}
    J(n) =\mu(n)-\alpha\Delta I(n)\label{eq:s7}
\end{equation}
which is the operation of a local mean minus a scaled local Laplacian of the signal.

To use this interpretation in the development of the edge-aware smoothing-sharpening filter, we
need to generalize the concept of Laplacian in the following sense.
In its original form, the Laplacian at a pixel location $n$ is defined
by \eqref{eq:laplacian} where the patch is centered at location $n$.
The Laplacian can also be written as
\begin{equation}
    \Delta I(n)=\frac{1}{N}\sum_{m\in\Omega_{n}}(I(m)-I(n))\label{eq:new laplacin}
\end{equation}
which is the average of the difference between the center
pixel $I(n)$ and each pixel $I(m)$ in the patch. Using this interpretation,
we make the following generalization. For any pixel $I(q)$ at location
$q\in\Omega_{n}$, the Laplacian is defined as
\begin{equation}
    \Delta I(q)=\frac{1}{N}\sum_{m\in\Omega_{n}}(I(m)-I(q))\label{eq:new laplacin-1}
\end{equation}
This definition of the Laplacian thus generalizes the concept from
the original one which only applies to the center pixel of the patch to
the one which applies to all pixels in the patch. Although the original
meaning of the Laplacian is lost in the generalization, we will use
the same name in this paper to simplify the terminology. Using this generalization,
the filter can be defined for all pixels in the patch as follows

\begin{align}
    J(q) & =\mu(n)-\alpha\Delta I(q)\label{eq:patch filter}\\
     & =\alpha I(q)+(1-\alpha)\mu(n)\label{eq:patchIntModel}
\end{align}
The difference between the two filters defined in \eqref{eq:s7} and
\eqref{eq:patch filter} is that the former is defined for the center pixel of the patch
while the latter is defined for all pixels in the patch. We will call
the new filter (stated in \eqref{eq:patchIntModel}) a patch interpolation
model, because the output is a weighted average of the input and the patch mean.

When the parameter $\alpha$ is fixed, the filter does not have an edge-awareness capability. This is a major problem of the filter. We will show in the next section that the idea of the guided filter provides a solution to this problem by adaptively setting the
parameter $\alpha$.

\subsection{The self-guided filter}

We revisit the basic idea of the guided filter  \cite{he2012guided}. A square patch of
radius $r$ has $N=(2r+1)^{2}$ pixels. Let $\Omega_{k}$
represent the $k$th patch in which a pixel at location $q$ is denoted
$I_{k}(q)$ where $q\in\Omega_{k}$ and the subscript $k$ indicates the patch. A linear model is imposed on
each pixel of the patch such that
\begin{equation}
    J_{k}(q)=a_{k}I_{k}(q)+b_{k},\label{eq:linear model}
\end{equation}
where $J_{k}(q)$ is the desired output. The two patch dependent parameters
$a_{k}$ and $b_{k}$ are determined by solving a regularized least
squares problem with the following cost function 
\begin{equation}
    c=\frac{1}{2N}\sum_{q\in\Omega_{k}}\left[J_{k}(q)-I_{k}(q)\right]^{2}+\frac{\epsilon}{2}a_{k}^{2},\label{eq:cost function C}
\end{equation}
where the second term is the regularization and $\epsilon$ is a user
defined parameter. Solving $\partial c/\partial b_{k}=0$, we have
$b_{k}=(1-a_{k})\mu_{k}$ where $\mu_{k}=\frac{1}{N}\sum_{q\in\Omega_{k}}I_{k}(q)$
is the patch mean. By Substitution of this result into \eqref{eq:linear model},
we have another patch model 
\begin{equation}
    J_{k}(q)=a_{k}I_{k}(q)+(1-a_{k})\mu_{k}\label{eq:GF patch model}
\end{equation}
which is an interpolation between the pixels in the patch and the
patch mean.

A new regularized least squares problem is obtained by substitution
of \eqref{eq:GF patch model} into \eqref{eq:cost function C} which
results in the following cost function
\begin{equation}
    d=\frac{\sigma_{k}^{2}}{2}(a_{k}-1)^{2}+\frac{\epsilon}{2}a_{k}^{2}\label{eq:cost function D}
\end{equation}
where $\sigma_{k}^{2}=\frac{1}{N}\sum_{q\in\Omega_{k}}\left[I_{k}(q)-\mu_{k}\right]^{2}$
is the patch variance. Solving $\partial d/\partial a_{k}=0$, we
have 
\begin{equation}
    a_{k}=\frac{\sigma_{k}^{2}}{\sigma_{k}^{2}+\epsilon}\label{eq:a_k}
\end{equation}

 Let $I(p)$
represent the pixel to be processed at location $p\in\Omega_k$. It can be shown that the pixel $I(p)$ belongs to $N$ overlapping patches \cite{he2012guided}. 
Since each patch model produces one output 
\begin{equation}
    J_{k}(p)=a_{k}I(p)+(1-a_{k})\mu_{k},\label{eq:J_k(p)}
\end{equation}
there are $N$ modelling results $\{J_{k}(p)\}_{k=1:N}$. To aggregate these results, a weighted average \cite{li2015weighted, Sun2020} is performed:
\begin{equation}
    J(p) =\sum_{k=1}^{N}w_{k}J_{k}(p) =I(p)\sum_{k=1}^{N}w_{k}a_{k}+\sum_{k=1}^{N}w_{k}(1-a_{k})\mu_{k}
    \label{eq:weighted average}
\end{equation}
where $\sum_{k=1}^{N}w_{k}=1$. The original guided filter \cite{he2012guided} uses the fixed weight $w_k=1/N$.

\subsection{Main idea of the proposed filter}

The key to our development is the patch interpolation model which
is stated in \eqref{eq:GF patch model} and the smooth-sharpening
filter stated in \eqref{eq:patchIntModel}. Comparing the two, we
can see that they are in a similar form. There are two key differences.
\begin{itemize}
\item In its general form \eqref{eq:smoothing}, the smoothing-sharpening filter
can be configured as either smoothing or sharpening. It is not obvious
how to set the parameter $\alpha$ such that the filter has the edge-aware
capability.
\item On the other hand, in the guided filter case, the filter parameter
$a_{k}$ is specifically determined for edge-awareness. Referring
to \eqref{eq:a_k}, for the case $\epsilon<<\sigma_{k}^{2}$ which
indicates strong texture/edge inside the patch, the algorithm assigns
$a_{k}\rightarrow1$ such that $J_{k}(q)\rightarrow I(q)$. For the
case $\epsilon>>\sigma_{k}^{2}$ which indicates a smooth patch, the
algorithm assigns $a_{k}\rightarrow0$ such that $J_{k}(q)\rightarrow\mu_{k}$.
As a result, edge-aware smoothing is achieved. However, it is always
a smoothing filter because $a_{k}=\frac{\sigma_{k}^{2}}{\sigma_{k}^{2}+\epsilon}<1$.
\end{itemize}
Based on the above observations, to develop an edge-aware smoothing-sharpening
filter we should find a generalization of the guided filter such that
it is possible to set $a_{k}\ge1$. We show in the next section that
the development is based on two key ideas:

\begin{enumerate}
\item using the patch interpolation
model stated in equation \eqref{eq:GF patch model}, and 

\item using the principle of maximum a posteriori to determine
the parameter $a_{k}$ to overcome the limitation of $a_{k}<1$ in
the original guided filter.
\end{enumerate}

\section{The edge-aware smoothing-sharpening filter\label{sec:The-proposed-smoothing-sharpenin}}
We first develop the self-guided edge-aware smoothing-sharpening filter in section \ref{sec:sg}. We then develop the external-guided version of the filter in section \ref{sec:eg}. Implementation and computational complexity of the proposed filter are discussed in section  \ref{sec:comp}.

\subsection{The self-guided form and its properties} \label{sec:sg}

\subsubsection{The self-guided smoothing-sharpening filter\label{subsec:The-self-guided-smoothing-sharpe}}

We define the linear Gaussian observation model for the patch data as follows
\begin{equation}
    I_{k}(q)=J_{k}(q)+r(q)\label{eq:ss2}
\end{equation}
where $r(q)$ is a realization of an i.i.d. zero mean Gaussian random
variable with variance $\tau^{2}N$ ($\tau>0$). Here we use the patch
size $N=|\Omega_{k}|$ to parameterize the noise variance such that
the result does not depend on $N$.

To determine $\alpha_{k}$, we treat it as a random variable and use
the principle of maximum a posteriori. More specifically, using Bayes rule we can write the negative log-posterior as a cost function
$D(\alpha_{k})$ by ignoring constants as follows
\begin{align}
    D(\alpha_{k}) & =-\log p(\alpha_{k}|\{I_{k}(q)\})\nonumber \\
                 & =-\log p(\{I_{k}(q)\}|\alpha_{k})-\log p(\alpha_{k})\label{eq:posterior}
\end{align}
where based on the observation model stated in \eqref{eq:ss2}, the negative log-likelihood for the patch is given by 
\begin{align}
    -\log p(\{I_{k}(q)\}|\alpha_{k}) & =\frac{1}{2\tau^{2}N}\sum_{q\in\Omega_{k}}(I_{k}(q)-J_{k}(q))^{2}\nonumber \\
                                    & =\frac{\sigma_{k}^{2}}{2\tau^{2}}(\alpha_{k}-1)^{2}\label{eq: patch loglikelihood}
\end{align}
Compared with the first term of the cost function of the guided filter stated in \eqref{eq:cost function C}, the above negative log-likelihood has an extra parameter $\tau$.
We set $\tau=1$ in this work which allows us to include the guided
filter as a special case in the proposed filter.

Substitution of \eqref{eq: patch loglikelihood} into \eqref{eq:posterior}
the cost function can be written as 
\begin{equation}
    D(\alpha_{k})=\frac{\sigma_{k}^{2}}{2}(\alpha_{k}-1)^{2}-\log p(\alpha_{k}).\label{eq:New D}
\end{equation}
Compared with the cost function of the guided filter, the above cost
function is different in two aspects: (a) the parameter $b_{k}$ is
implicitly defined in the patch interpolation model, and (b) the regularization
term $\frac{\epsilon}{2}\alpha_{k}^{2}$ (which is the negative of
logarithm of zero mean Gaussian) is replaced by the negative log-prior
which permits us to develop different filters.

In this work, we consider the generalized Gamma distribution as the
prior: 
\begin{equation}
    p(\alpha_{k})\propto\alpha_{k}^{\eta}e^{-\frac{1}{2}(\alpha_{k}/\theta)^{h}}\label{eq:R}
\end{equation}
where $\theta>0$ is a scale parameter. We set $h=2$
and $\eta\ge0$ to control the shape of the distribution. Substituting
\eqref{eq:R} into \eqref{eq:New D} we have the cost function in
which constant terms are omitted
\begin{equation}
    D(\alpha_k)=\frac{\sigma_{k}^{2}}{2}(\alpha_{k}-1)^{2}+\frac{1}{2\theta^{2}}\alpha_{k}^{2}-\eta\log\alpha_{k}\label{eq: NEW D}
\end{equation}
We can easily see the motivation and justification of such settings.
When $\eta=0$, the cost function is the same as that of the guided filter
with the setting $\epsilon=1/\theta^{2}$. When $\eta>0$, the cost
function has an extra term $-\eta\log\alpha_{k}$ compared with the cost
function of the guided filter. We will show that this extra term permits
the filter to be configured as either a smoothing filter ($0<\alpha_{k}\le1$)
or a sharpening filter ($\alpha_{k}>1$). 

Another justification is mathematical simplicity. The cost function under this parameter setting is convex leading to a unique minimum. Indeed, the generalized Gamma
distribution allows us to explore other settings of parameters such
as $h\ne2$. However, for such a setting, the cost function may not be convex and is difficult to optimize. Therefore, we do not pursue study in this direction.

We now determine the filter parameter $\alpha_k$ by minimizing the cost function $D(\alpha_k)$ which is equivalent to maximising the posterior. Solving $\partial D/\partial\alpha_{k}=0$, we obtain the optimal
value
\begin{equation}
    \alpha_{k}=\frac{1}{2}\left\{ \frac{\sigma_{k}^{2}}{\sigma_{k}^{2}+\frac{1}{\theta^{2}}}+\sqrt{\left(\frac{\sigma_{k}^{2}}{\sigma_{k}^{2}+\frac{1}{\theta^{2}}}\right)^{2}+\frac{4\eta}{\sigma_{k}^{2}+\frac{1}{\theta^{2}}}}\right\} \label{eq:alpha_k new}
\end{equation}

\subsubsection{Properties} \label{sec:prop}

To reveal how the proposed filter generalizes the original guided
filter, we set $\epsilon=1/\theta^{2}$ such that the guided filter's
parameter stated in \eqref{eq:a_k} is given by
\begin{equation}
    a_{k}=\frac{\sigma_{k}^{2}}{\sigma_{k}^{2}+\epsilon}=\frac{\sigma_{k}^{2}}{\sigma_{k}^{2}+1/\theta^{2}}\label{a_k new}
\end{equation}
Substitution \eqref{a_k new} into \eqref{eq:alpha_k new}, we have
\begin{equation}
    \alpha_{k}=\frac{1}{2}\left\{ a_{k}+\sqrt{a_{k}^{2}+\frac{4\eta}{\sigma_{k}^{2}+\epsilon}}\right\} \label{eq:slef-guided alpha_k}
\end{equation}
We can clearly see that the parameter of the proposed filter can be expressed as a function of the parameter of the original guided filter. Their relationship is discussed in the following.
Based on the interpolation model stated in equation \eqref{eq:GF patch model}, 
we can prove that the proposed filter can be configured by setting
$\eta$ relative to $\epsilon$ as follows.
\begin{itemize}
\item When $\eta<\epsilon$, we can show that $\alpha_{k}<1$ which leads to a smoothing
filter. In an extreme case when $\eta=0$, the proposed filter is
reduced to the original guided filter. It is also interesting to note
that for this setting $\alpha_{k}>a_{k}$ which means the proposed
filter always performs a lower degree of smoothing than the original
guided filter. 
\item When $\eta=\epsilon$, we can show that $\alpha_{k}=1$ which leads to no filtering.
\item When $\eta>\epsilon$, we can show that $\alpha_{k}>1$ which leads to a sharpening
filtering. The sharpening gain is defined as $\gamma_{k}=\alpha_{k}-1$
which is an increasing function of $\eta$.
\end{itemize}
In light of the above discussion and to simplify parameter settings,
we introduce another parameter $\kappa$ to replace $\eta$ by defining
$\eta=\kappa/\theta^{2}=\kappa\epsilon.$ The parameter of the proposed
filter can be re-written as
\begin{equation}
    \alpha_{k}=\frac{1}{2}\left\{ a_{k}+\sqrt{a_{k}^{2}+4\kappa(1-a_{k})}\right\} \label{eq:self guided alpha_k}
\end{equation}
The advantage of this new parameterization is that the filter can
be configured as smoothing by setting $0\le\kappa<1$ and as sharpening
by setting $\kappa>1$. 

\begin{figure}
    \begin{centering}
    \includegraphics[scale=0.75]{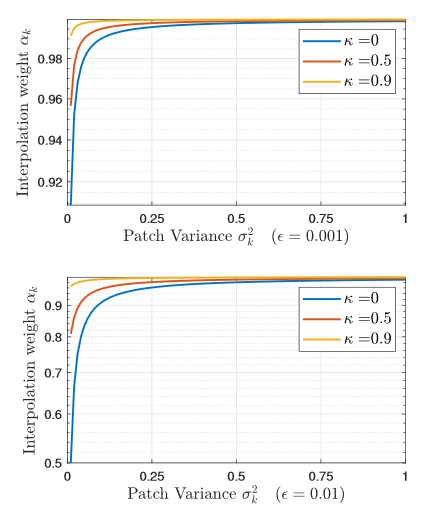}
    \par\end{centering}
    \caption{\label{fig:Interpolation-weight-}Interpolation weight $\alpha_{k}$
as a function of patch variance $\sigma_{k}^{2}$ and filter parameter
$\kappa$ under two fixed settings of $\epsilon$
(top $\epsilon=0.001$, bottom $\epsilon=0.01$). As $\kappa$ is increased, $\alpha_{k}$ is less adapted
to $\sigma_{k}^{2}$ and is close to the constant 1 for $\sigma_{k}^{2}>T$
where $T$ is a signal dependent threshold.}

\end{figure}

For the smoothing case, we plot $\alpha_{k}$
for different settings of $\kappa$ including the case with $\kappa=0$
which is the guided filter. Results are shown in \figref{fig:Interpolation-weight-}.
We can see that as $\kappa$ is increased, $\alpha_{k}$ is less adapted
to $\sigma_{k}^{2}$ and is closer to the constant 1 (no filtering) for $\sigma_{k}^{2}>T$
where $T$ is a signal dependent threshold. 

\begin{figure}
    \begin{centering}
    \includegraphics[scale=0.75]{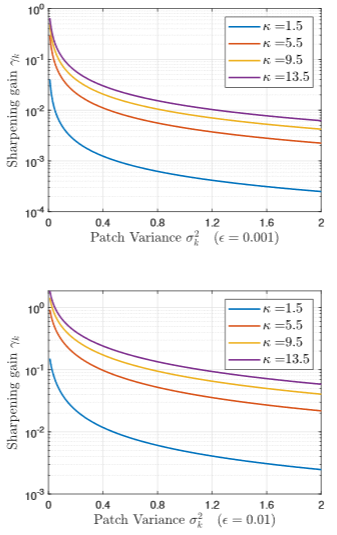}
    \par\end{centering}
    \caption{\label{fig:The-sharpening-gain}The sharpening gain $\gamma_{k}$
is a decreasing function of the patch variance $\sigma_{k}^{2}$ for
various settings of $\kappa$ under two fixed settings of $\epsilon$
(top $\epsilon=0.001$, bottom $\epsilon=0.01$). The sharpening gain
$\gamma_{k}$ is bigger for a bigger value of $\kappa$ and is a decreasing
function of the patch variance $\sigma_{k}^{2}$.}
\end{figure}

Next, we study the sharpening gain $\gamma_k$ as a function of $\kappa$ ($\kappa>1$)
and the patch variance $\sigma_{k}^{2}$. The interpolation model
can be re-written in a sharpening filter form
\begin{equation}
    J_{k}(q)=I_{k}(q)+\gamma_{k}(I_{k}(q)-\mu_{k})\label{eq:sharpening filter}
\end{equation}
\figref{fig:The-sharpening-gain} shows the sharpening gain as
a function of the patch variance $\sigma_{k}^{2}$ for various settings
of $\kappa$. We can make the following observations. (a) The sharpening
gain is a decreasing function of the patch variance. This is a desirable
property of the sharpening filter. It performs a higher degree of
enhancement on a patch of smaller variance which is usually due to
low contrast. (b) The level of enhancement can be controlled by setting
the parameter $\kappa$. A bigger value will lead to a bigger sharpening
gain for the patch of fixed variance. (c) The parameter $\epsilon$
also controls the sharpening gain. For the same setting of $\kappa$,
a bigger value of $\epsilon$ will lead to a bigger value of the sharpening
gain.

\subsection{The guided form and its properties} \label{sec:eg}

In this section, we develop a guided version of the proposed
filter by a further generalization of smoothing-sharpening formulation
stated in section (\ref{subsec:The-smoothing-sharpening-filter})
and the guided filter.

\subsubsection{The guided smoothing-sharpening formulation and the guided filter}

Referring to the filter formulation stated in \eqref{eq:patch filter},
we can see that one of the key components is the Laplacian which is
a second derivative operation and is thus sensitive to noise in the
image $I$. When a guidance image $G$, which is assumed to have a
higher signal-to-noise ratio, is available, a generalization is to
replace the Laplacian calculated on $I$ by the Laplacian calculated
on $G$ such that the filter can be written as:

\begin{equation}
    J(q)=\mu(n)-\alpha\Delta G(q)\label{eq:g ss model}
\end{equation}
where $\mu(n)$ is the mean of the patch centered at $I(n)$ and 
\begin{equation}
    \Delta G(q)=\frac{1}{N}\sum_{m\in\Omega_{n}}(G(m)-G(q))\label{eq:g laplacian}
\end{equation}
Although this idea is technically sound, how to determine $\alpha$
remains a problem. We will revisit the basic idea of the original
guided filter and show how this problem can be solved by using a similar
approach as the development of the self-guided version of the proposed
filter described in the previous section.

In the original guided filter, the patch model is given by
\begin{equation}
    J_{k}(q)=a_{k}G_{k}(q)+b_{k}\label{eq:guided patch model}
\end{equation}
The two parameters $a_{k}$ and $b_{k}$ are determined by minimization
of the cost function over the patch data 
\begin{equation}
    \{a_{k},b_{k}\}=\min_{a_{k},b_{k}}\sum_{q\in\Omega_{k}}\left[J_{k}(q)-I_{k}(q)\right]^{2}+\frac{\epsilon}{2}a_{k}^{2}\label{eq:cost function}
\end{equation}
It can be shown that 
\begin{equation}
    b_{k}=\mu_{k}-a_{k}\nu_{k}\label{eq:guided b_k}
\end{equation}
where $\mu_{k}$ and $\nu_{k}$ are the mean for the $k$th patch
of image $I$ and $G$, respectively. The parameter $a_{k}$ is
given by
\begin{equation}
    a_{k}=\frac{\phi_{k}}{\varsigma_{k}^{2}+\epsilon}\label{eq:guided ak}
\end{equation}
where $\varsigma_{k}^{2}$ is the patch variance of $G$, and $\phi_{k}$
is the sample covariance
\begin{equation}
    \phi_{k}=\frac{1}{N}\sum_{q\in\Omega_{k}}(G_{k}(q)-\nu_{k})(I_{k}(q)-\mu_{k})\label{eq:guided cov}
\end{equation}
Substituting \eqref{eq:guided b_k} into \eqref{eq:guided patch model},
we have a new patch model
\begin{equation}
    J_{k}(q) =\mu_{k}+a_{k}(G_{k}(q)-\nu_{k}) =\mu_{k}-a_{k}\Delta G_{k}(q)\label{eq:new guided patch model}
\end{equation}
where $\Delta G_{k}(q)=\nu_{k}-G_{k}(q)$ is the generalization of the 
Laplacian defined in \eqref{eq:new laplacin-1}.

Comparing the guided filter model stated in \eqref{eq:new guided patch model}
with the proposed smoothing-sharpening model stated in \eqref{eq:g ss model},
we can see that they are similar. Simply put, if we start with the
filter model stated in \eqref{eq:g ss model}, we will determine the
same filter as the guided filter which is a smoothing filter. In the next
section, we address the problem of how to develop a smoothing-sharpening
guided filter.

\subsubsection{The proposed guided smoothing-sharpening filter}

To use \eqref{eq:new guided patch model} to develop the smoothing
and sharpening filter, we face a new difficulty that $a_{k}$ can
be positive or negative. This is unlike the case for the self-guidance version 
developed in section \ref{subsec:The-self-guided-smoothing-sharpe} where it is always the case $a_{k}\ge0.$
Referring to \eqref{eq:guided ak}, the sign of $a_{k}$ is defined
by the sign of the covariance $\phi_{k}$ such that
\begin{equation}
    \text{sign}(a_{k})=\text{sign}(\phi_{k})\label{eq:sign}
\end{equation}
The role of the sign of the covariance $\phi_{k}$ can be explained
as follows. Because $\Delta I_{k}(q)$ is replaced by $\Delta G_{k}(q)$,
it requires that the two patches $I_{k}$ and $G_{k}$ must be correlated.
If they are positively correlated, then it is expected the two Laplacians
are of the same sign. If they are negatively correlated, then it is
expected the two Laplacians are of opposite sign. Thus a correction
of the sign of the Laplacian calculated on the guidance image is required.
The definition of $a_{k}$ stated in \eqref{eq:guided ak} automatically
satisfies this requirement.

In light of the above discussion, we can develop the guided version
of the smoothing-sharpening filter by defining the guided patch interpolation
model as the following
\begin{equation}
    J_{k}(q)=\mu_{k}+\text{sign}(\phi_{k})\alpha_{k}(G_{k}(q)-\nu_{k})\label{eq:new new guided patch model}
\end{equation}
such that $\alpha_{k}$ is a positive parameter. We can then follow
the same procedure as that presented in section \ref{subsec:The-self-guided-smoothing-sharpe}
for the development. More specifically, the negative
log-likelihood is
\begin{align}
    -\log p(\{J_{k}(q)\}|\alpha_{k}) & =\frac{1}{2\tau^{2}N}\sum_{q\in\Omega_{k}}(I_{k}(q)-J_{k}(q))^{2}\nonumber \\
     & =\frac{\varsigma_{k}^{2}}{2}\alpha_{k}^{2}-|\phi_{k}|\alpha_{k}\label{eq:patach loglikelihood-1}
\end{align}
where we have set $\tau=1$ as before. Using the same generalized
Gamma distribution as the prior, we obtain the negative log-posterior
as the cost function $D(\alpha_{k})$
\begin{align}
    D(\alpha_{k}) & =-\log p(\{J_{k}(q)\}|\alpha_{k})-\log p(\alpha_{k})\nonumber \\
     & =\frac{\varsigma_{k}^{2}}{2}\alpha_{k}^{2}-|\phi_{k}|\alpha_{k}+\frac{1}{2\theta^{2}}\alpha_{k}^{2}-\eta\log\alpha_{k}\label{eq:guided posterior}
\end{align}
We also follow the same parameter settings of the self-guided form
by letting $\epsilon=1/\theta^{2}$ and $\eta=\kappa\epsilon$. Solving
$\partial D/\partial\alpha_{k}=0$ and re-arranging the results, we
have 
\begin{equation}
    \alpha_{k}=\frac{1}{2}\left\{ \frac{|\phi_{k}|}{\varsigma_{k}^{2}+\epsilon}+\sqrt{\left(\frac{|\phi_{k}|}{\varsigma_{k}^{2}+\epsilon}\right)^{2}+\frac{4\kappa\epsilon}{\varsigma_{k}^{2}+\epsilon}}\right\} \label{eq:alpha_k}
\end{equation}

\subsubsection{Properties}

We can clearly see that when $G=I,$ we have $\varsigma_{k}^{2}=\phi_{k}=\sigma_{k}^{2}$.
The guided version of the proposed filter stated by \eqref{eq:guided alpha_k}
reduces to its self-guided version stated by \eqref{eq:slef-guided alpha_k}.
The relationship between the proposed filter and original guided filter
can be revealed by substitution of \eqref{eq:guided ak} into \eqref{eq:alpha_k},
which results in the following
\begin{equation}
    \alpha_{k}=\frac{1}{2}\left\{ |a_{k}|+\sqrt{a_{k}^{2}+\frac{4\kappa\epsilon}{\varsigma_{k}^{2}+\epsilon}}\right\} \label{eq:guided alpha_k}
\end{equation}
We can also see that $\alpha_{k}\ge|a_{k}|$ from \eqref{eq:guided alpha_k}.
The original guided filter is a special case of the proposed filter
when $\kappa=0$ leading to $\alpha_{k}=|a_{k}|$.

Next we discuss under what parameter setting the filter is smoothing
or sharpening. The analysis presented in section \ref{subsec:The-smoothing-sharpening-filter}
can not be used, because the Laplacian is calculated on the guidance
image rather than on the image to be processed. The following analysis is based on an observation that a smoothing filter will
reduce the patch variance while a sharpening filter will increase
the patch variance. It is also assumed that the patch mean is not changed
by the filter. This is a reasonable assumption because smoothing and
sharpening usually do not change the average brightness of the image.

To make the discussion easy to follow, we first define the patch variance for the three images as
\begin{equation}
\text{Original image} : \sigma_{k}^{2}=\frac{1}{N}\sum_{q\in\Omega_{k}}(I_{k}(q)-\mu_{k})^{2}\label{eq:sigma2}
\end{equation}

\begin{equation}
\text{Guided image} : \varsigma_{k}^{2}=\frac{1}{N}\sum_{q\in\Omega_{k}}(G_{k}(q)-\nu_{k})^{2}\label{eq:varsigma2}
\end{equation}

\begin{equation}
\text{Filtered image} :\tau_{k}^{2}=\frac{1}{N}\sum_{q\in\Omega_{k}}(J_{k}(q)-\mu_{k})^{2}\label{eq:tau}
\end{equation}
In equation \eqref{eq:tau}, we substitute $J_k(q)$  by the patch interpolation model stated in \eqref{eq:new new guided patch model} and use the definition stated in \eqref{eq:varsigma2}, we obtain \begin{equation}
\tau_{k}^{2} = \alpha_k^2 \varsigma_k^2\label{eq:tau3}
\end{equation}

Next we identify parameter settings that lead to reduced variance, i.e., $\tau_{k}^{2}/\sigma_{k}^{2}\le1$ for smoothing or increased variance, i.e., 
$\tau_{k}^{2}/\sigma_{k}^{2}>1$ for sharpening. It can be shown that
\begin{equation}
    \frac{\tau_{k}^{2}}{\sigma_{k}^{2}}  =\frac{1}{2}\rho_{k}^{2}\hat{\varsigma}_k^2\left\{ 1+\sqrt{1+\frac{4\kappa\epsilon}{\sigma_{k}^{2}\rho_{k}^{2}\hat{\varsigma}_k^2}}\right\} +\frac{\kappa\epsilon\hat{\varsigma}_k^2}{\sigma_{k}^{2}}\label{eq:rati0}
\end{equation}
where $\hat{\varsigma}_k^2 = \frac{\varsigma_{k}^{2}}{\varsigma_{k}^{2}+\epsilon}$ and $\rho_{k}$ ($|\rho_{k}|\le1$) is the cross correlation coefficient of the two patches defined as 
\begin{equation}
    \rho_{k}=\frac{\phi_{k}}{\sigma_{k}\varsigma_{k}}
\end{equation}
where $\phi_k$ is the sample covariance between the corresponding patches of the original and guided images. It is defined in equation \eqref{eq:guided cov}.

In a special case in which $\kappa=0$ leading to the original guided
filter, we have the following results
\begin{equation}
    \frac{\tau_{k}^{2}}{\sigma_{k}^{2}}=\rho_{k}^{2}\eta_{k}^{2}=\rho_{k}^{2}\left[\frac{\varsigma_{k}^{2}}{\varsigma_{k}^{2}+\epsilon}\right]^{2}<1\label{guided filter smooth}
\end{equation}
This means that the original guided filter is always a smoothing filter.

However, there is a highly non-linear relationship between the ratio
$\tau_{k}^{2}/\sigma_{k}^{2}$ and the two filter parameters $\kappa$
and $\epsilon$ for the two  patches in $I$ and $G$ where
the variances ($\sigma_{k}^{2}$ and $\varsigma_{k}^{2}$) and correlation
coefficients ($\rho_{k}$) can be calculated. In theory, we can calculate
the required parameter settings such that for two given patches the
ratio is greater than one or less than one. But doing so adds considerable
computation burden in practice. A practical approach is to let the
user set the two parameters such that the desirable result is produced.
In this regard, we can see that the ratio $\tau_{k}^{2}/\sigma_{k}^{2}$
is an increasing function of $\frac{\kappa\epsilon}{\sigma_{k}^{2}}$.
For the case where $\epsilon$ and $\sigma_{k}^{2}$ are fixed, increasing
$\kappa$ will increase the ratio and the patch is more likely to
be sharpened (or less smoothed). For the case where $\kappa$ and
$\epsilon$ are fixed, a patch with a larger/smaller value of variance
is less/more likely to be sharpened. This is a desirable feature because
more sharpening should be applied to areas of less variance.

\begin{figure*}
    \centering
\begin{subfigure}{0.19\linewidth}
    \includegraphics[width=\linewidth]{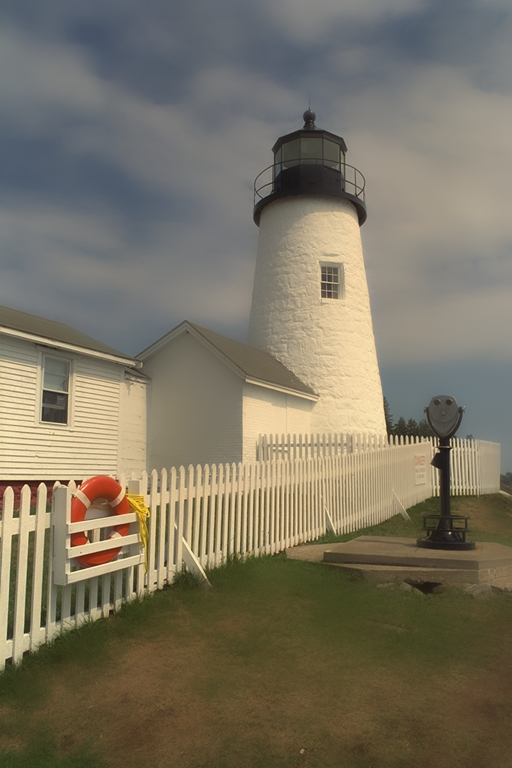}
    \subcaption{$\kappa =0.01 $}
    \label{fig:ps_kappa_a}
\end{subfigure}
\begin{subfigure}{0.19\linewidth}
    \includegraphics[width=\linewidth]{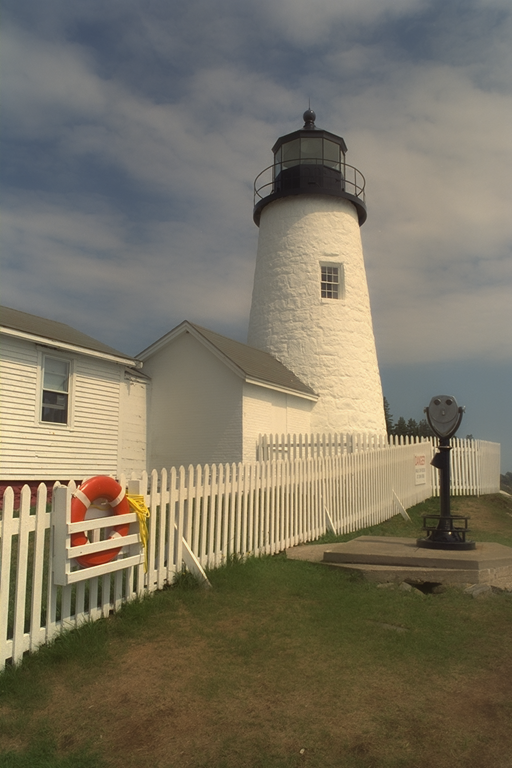}
    \subcaption{$\kappa = 0.1$}
    \label{fig:ps_kappa_b}
\end{subfigure}
\begin{subfigure}{0.19\linewidth}
    \includegraphics[width=\linewidth]{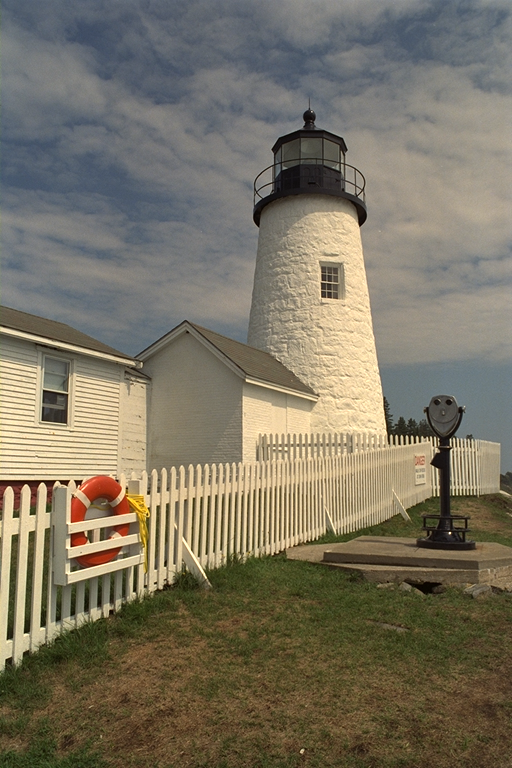}
    \subcaption{$\kappa = 1$}
        \label{fig:ps_kappa_c}

\end{subfigure}
\begin{subfigure}{0.19\linewidth}
    \includegraphics[width=\linewidth]{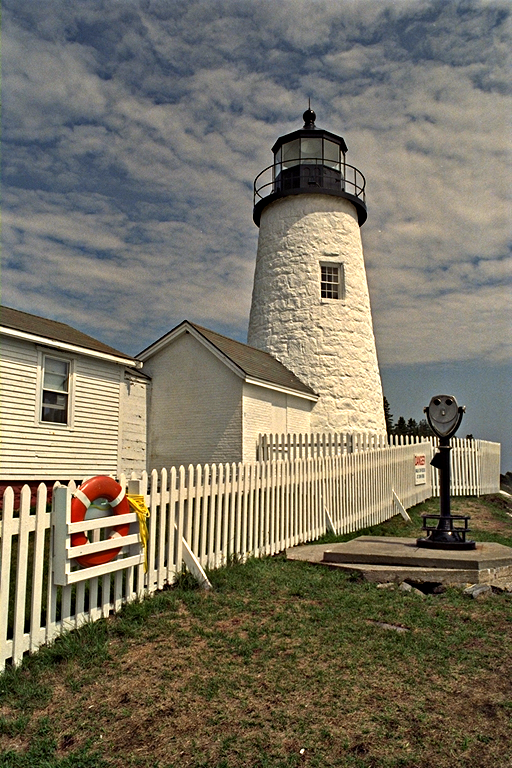}
    \subcaption{$\kappa = 5$}
        \label{fig:ps_kappa_d}

\end{subfigure}
\begin{subfigure}{0.19\linewidth}
    \includegraphics[width=\linewidth]{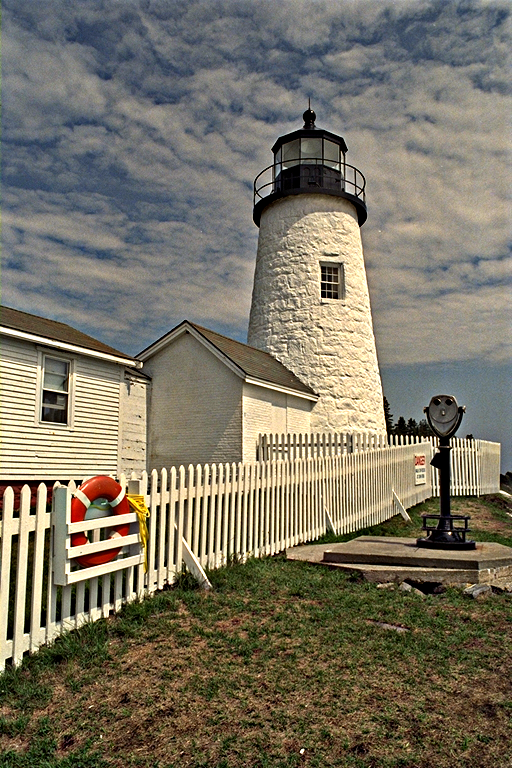}
    \subcaption{$\kappa = 7.5$}
        \label{fig:ps_kappa_e}

\end{subfigure}
\caption{Effect of varying $\kappa$ for settings $ r =11 , \epsilon = 0.01, scale = 1$. (a) and (b) Smoothing ($0< \kappa < 1$). (c) Original image ($\kappa = 1$, no filtering). (d) and (e) Sharpening ($\kappa >1$).}
\label{fig:ps_kappa}
\end{figure*}

\subsection{Implementation and computational complexity} \label{sec:comp}

The implementation is similar to that of the original guided filter.
For each patch, we calculate the parameter $\alpha_{k}$ using either
\eqref{eq:self guided alpha_k} for the self-guided version or \eqref{eq:alpha_k}
for the guided version. We then perform the weighted average operation.
More specifically, for the self-guided version, the filter output is
given by
\begin{equation}
    J(p)=I(p)\sum_{k=1}^{N}w_{k}\alpha_{k}+\sum_{k=1}^{N}w_{k}(1-\alpha_{k})\mu_{k}\label{eq:ave1}
\end{equation}
For the guided version, the filter output is given by
\begin{equation}
    J(p)=G(p)\sum_{k=1}^{N}w_{k}\beta_{k}+\sum_{k=1}^{N}w_{k}(\mu_{k}-\beta_{k}\nu_{k})\label{eq:ave3}
\end{equation} 
where $\beta_{k} =\text{sign}(\phi_{k})\alpha_{k}$ and $a_{k}$ is given by \eqref{eq:guided ak}. 

We calculate the weight
for the self-guided and the guided case as follows
\begin{equation}
    w_{k}=\frac{c_{k}}{1+\left(\sigma_{k}^{2}/(s\bar{\sigma}^{2})\right)^{2}}\label{eq:w1}
\end{equation}
and

\begin{equation}
    w_{k}=\frac{c_{k}}{1+\left(\varsigma_{k}^{2}/(s\bar{\varsigma}^{2})\right)^{2}}\label{eq:w2}
\end{equation}
where $c_{k}$ is a normalization factor to ensure $\sum_{k=1}^{N}w_{k}=1$,
$s$ is a user defined scale parameter, $\bar{\sigma}^{2}$ and $\bar{\varsigma}^{2}$
are the average of $\sigma_{k}^{2}$ and $\varsigma_{k}^{2}$ over
the whole image.

The proposed guided smoothing-sharpening filter can be implemented in MATLAB of which the code is shown in Appendix. The self-guided version has a similar implementation. We assume the four parameters are given:
patch radius ($r$), Kappa ($\kappa$), Epsilon $(\epsilon)$, and Scale $(s)$.
We can see from the brute-force implementation that the
proposed filter has an $\mathcal{O}(N)$ complexity which is the same
as that of the original guided filter. It can be implemented by using
7 linear filters. 

For color images, we can process each color component individually. Alternatively, we can  convert the image from RGB to HSV color
space. Filtering is performed on the value channel. The processing result is then combined with the hue and saturation channels and is converted back to RGB.

\section{Applications examples}\label{App}
There are two purposes of this section: validation (section \ref{sect:validate}) of the theoretical analysis of the proposed filter, and demonstration of successful applications in (a) adaptive smoothing and sharpening (section \ref{sec:ss}) based on extracted information of texture, depth and blurriness, and (b) information fusion (section \ref{sec:ext guided filt}) for denoising and creating high resolution multi-spectral images.  

\subsection{Effects of parameter settings for the self-guidance case} \label{sect:validate}
We demonstrate the properties of the filter and confirm the theoretical analysis presented in section \ref{sec:prop}. 
The proposed filter has 3 user defined parameters: (a) The patch radius $r$, (b) the sharpening/smoothing gain $\kappa$, and  (c) the regularization parameter $\epsilon$. In this section we study the effect of these parameters on the processed image.

In Fig. \ref{fig:ps_kappa} we demonstrate the effect of varying $\kappa$ by keeping the rest of the parameters fixed. Fig. \ref{fig:ps_kappa_a} and \ref{fig:ps_kappa_b} show the effect of smoothing when $0<\kappa < 1$. Smaller $\kappa$ values increase the smoothing level on the result image. Fig. \ref{fig:ps_kappa_d} and \ref{fig:ps_kappa_e} show the effect of sharpening when $\kappa > 1$. Larger $\kappa$ values produce a sharper result. The image shown in Fig. \ref{fig:ps_kappa_c} is produced by setting $\kappa = 1$. We can verify that it is exactly as the original image. Thus when $\kappa = 1$ the filter produces no smoothing or sharpening effect.

\begin{figure}[t]
    \centering
\begin{subfigure}{0.32\columnwidth}
    \includegraphics[width=\linewidth]{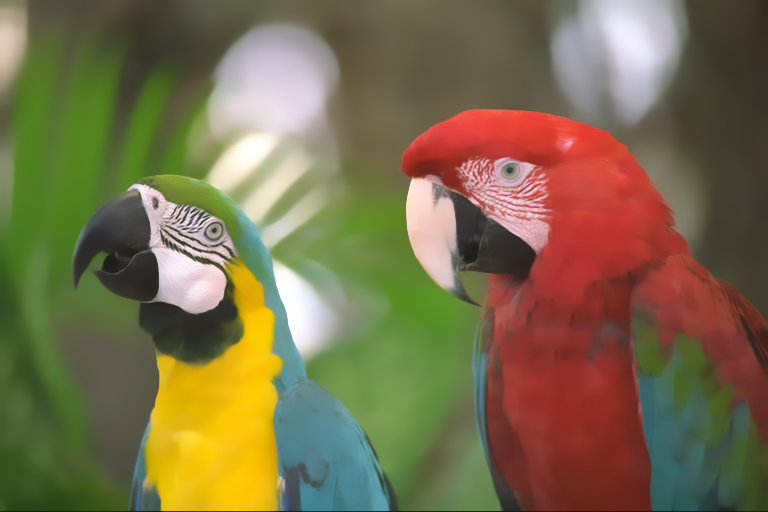}
    \subcaption{$r = 5, \epsilon =10^{-2} $}
\end{subfigure}
\begin{subfigure}{0.32\columnwidth}
    \includegraphics[width=\linewidth]{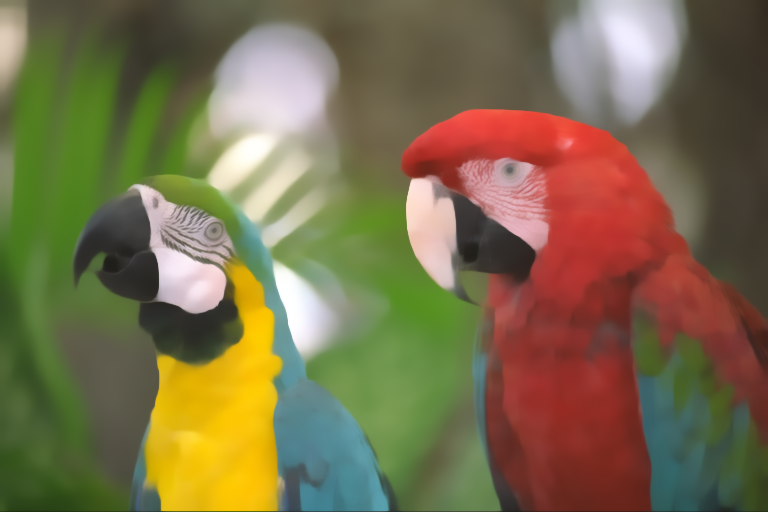}
    \subcaption{$r = 5, \epsilon = 10^{-1} $}
\end{subfigure}
\begin{subfigure}{0.32\columnwidth}
    \includegraphics[width=\linewidth]{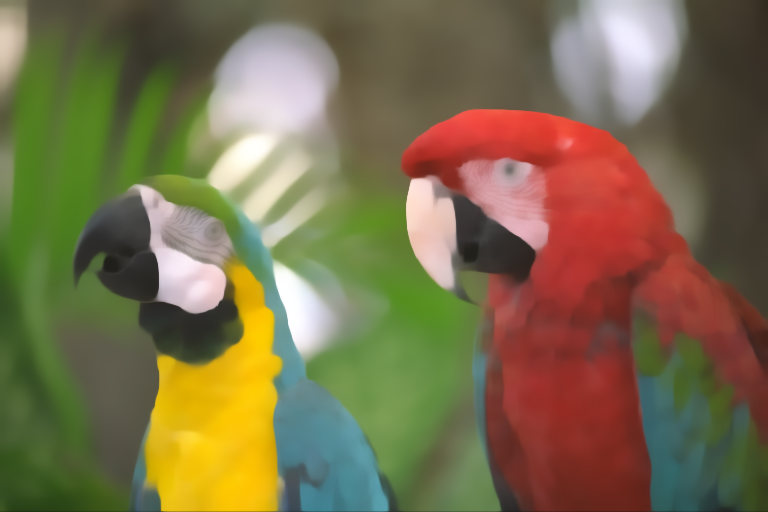}
    \subcaption{$r = 5, \epsilon = 1$}
\end{subfigure}
\begin{subfigure}{0.32\columnwidth}
    \includegraphics[width=\linewidth]{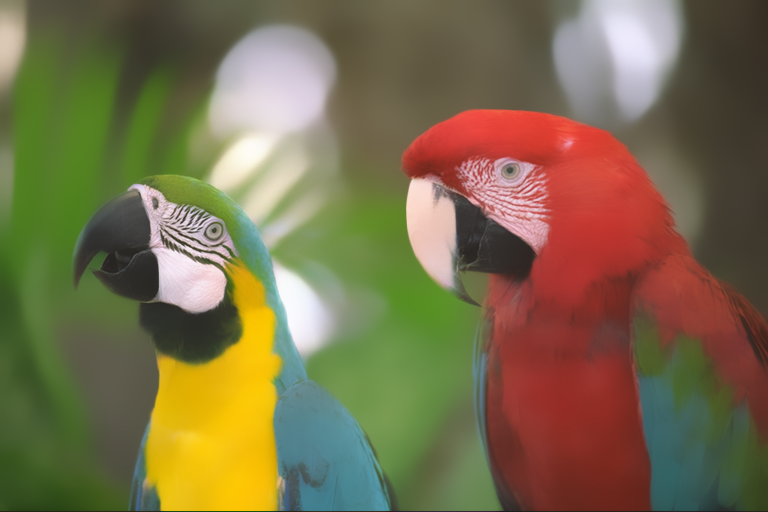}
    \subcaption{$r = 10, \epsilon = 10^{-2} $}
\end{subfigure}
\begin{subfigure}{0.32\columnwidth}
    \includegraphics[width=\linewidth]{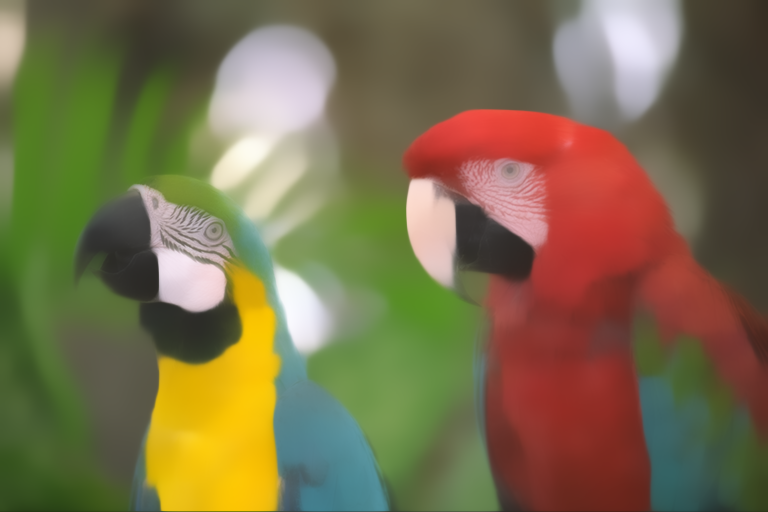}
    \subcaption{$r = 10, \epsilon =10^{-1}  $}
\end{subfigure}
\begin{subfigure}{0.32\columnwidth}
    \includegraphics[width=\linewidth]{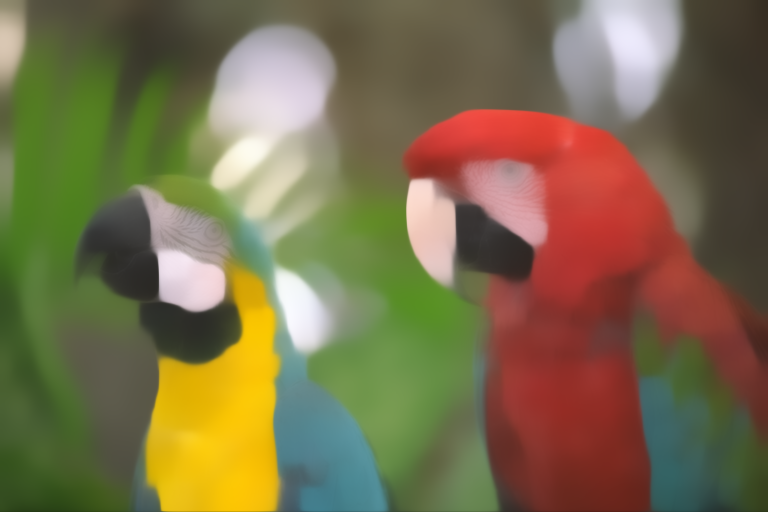}
    \subcaption{$r = 10, \epsilon = 1$}
\end{subfigure}
\caption{Effect of varying $\epsilon$ and $r$ in smoothing mode for fixed settings of  $ \kappa = 10^{-2}$ and $scale = 0.25$.}
\label{fig:PS_epsilon_smoothing}
\end{figure}

Next, we study the effect of patch size and $\epsilon$ for sharpening and smoothing separately. To set the filter in smoothing mode we set a fixed $\kappa = 10^{-2}$. In Fig. \ref{fig:PS_epsilon_smoothing} the results are organized in such a way that the radius varies from 5 to 10 column-wise while $\epsilon$ varies from $10^{-2}$ to $1$ row-wise. Results shown in this figure clearly show the edge preserving capabilities of the filter in smooth mode. Setting a larger $\epsilon$ value produces a more washed out result, while increasing the radius of the filter also produces a stronger smoothing result.

\begin{figure*}[h!]
    \centering
\begin{subfigure}{0.32\columnwidth}
    \includegraphics[width=\linewidth]{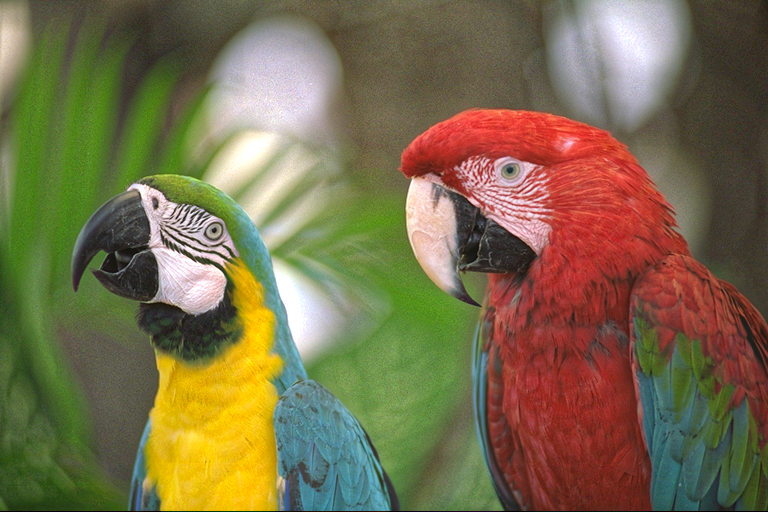}
    \subcaption{$r = 5, \epsilon =10^{-4} $}
\end{subfigure}
\begin{subfigure}{0.32\columnwidth}
    \includegraphics[width=\linewidth]{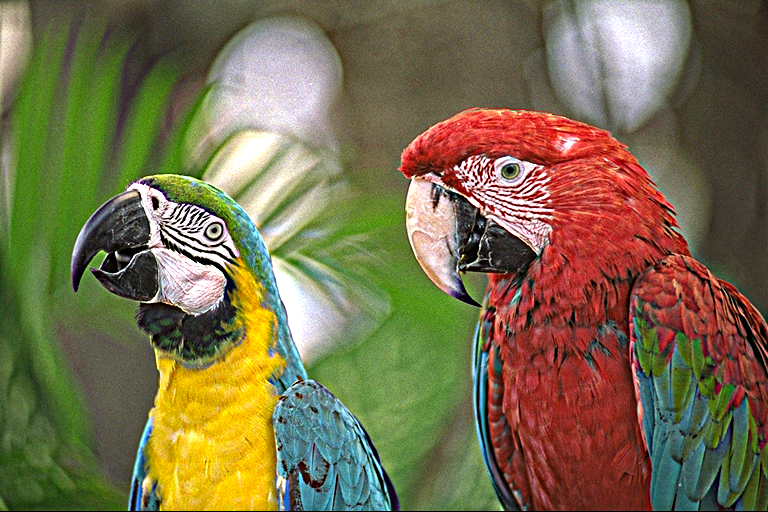}
    \subcaption{$r = 5, \epsilon =10^{-2} $}
\end{subfigure}
\begin{subfigure}{0.32\columnwidth}
    \includegraphics[width=\linewidth]{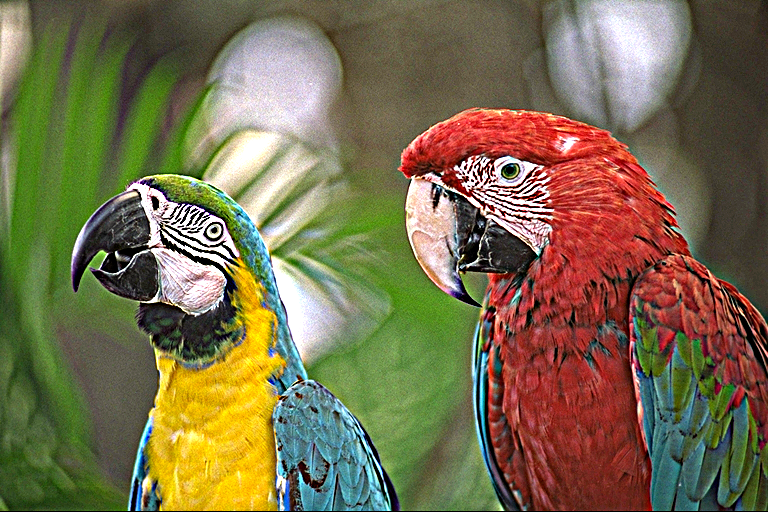}
    \subcaption{$r = 5, \epsilon =100 $}
\end{subfigure}

\begin{subfigure}{0.32\columnwidth}
    \includegraphics[width=\linewidth]{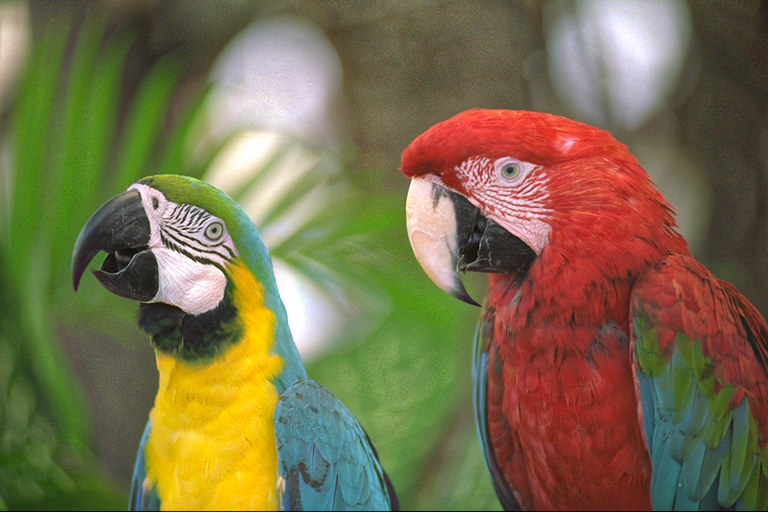}
    \subcaption{$r = 10, \epsilon = 10^{-4} $}
\end{subfigure}
\begin{subfigure}{0.32\columnwidth}
    \includegraphics[width=\linewidth]{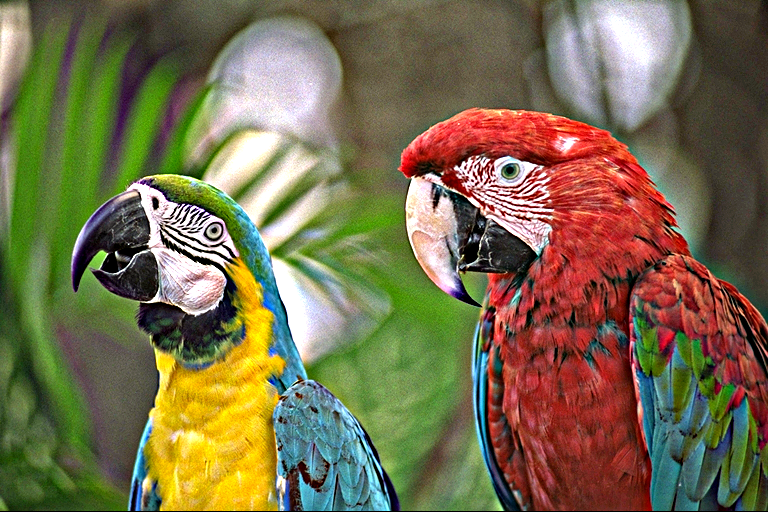}
    \subcaption{$r = 10, \epsilon = 10^{-2} $}
\end{subfigure}
\begin{subfigure}{0.32\columnwidth}
    \includegraphics[width=\linewidth]{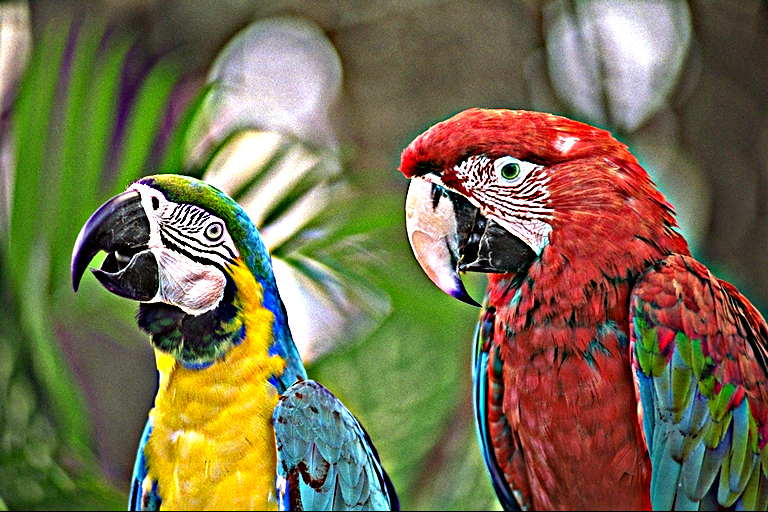}
    \subcaption{$r = 10, \epsilon = 100 $}
\end{subfigure}
    \caption{Effect of varying $\epsilon$ and $r$ in sharpening mode for fixed settings of $ \kappa = 20$ and $scale = 1$.}
    \label{fig:PS_epsilon_sharpenning}
\end{figure*}

The results shown in Fig. \ref{fig:PS_epsilon_sharpenning} are obtained by setting $\kappa = 20$ to demonstrate the sharpening effects as a function of  different $\epsilon$ and $r$ values. These results confirm that the sharpening gain increases with the increase in $\epsilon$ as previously stated in Fig. \ref{fig:The-sharpening-gain}. Increasing the radius impacts the variance for each pixel position. As a result, it produces a change in the sharpening gain.

\subsection{Application in adaptive smoothing-sharpening} \label{sec:ss}

As mentioned in the previous section, the value of $\kappa$ controls the sharpening/smoothing gain of the proposed filter, in this section we propose three pixel-adaptive smoothing and sharpening algorithms by defining $\kappa$  as a non-linear transformation of a feature map such as depth, blurriness or texture. 

The non-linear transformation used in this paper is simply a variation of the Gompertz\footnote{\href{https://en.wikipedia.org/wiki/Gompertz_function}{https://en.wikipedia.org/wiki/Gompertz\_function}} function which is a sigmoid function. It is defined as:
 
  \begin{equation}
     g(t) =  ae^{-be^{-c t}}\label{eq:gompertz}
 \end{equation}
 where $a$, $b$ and $c$ are three parameters.
Let us consider $t$ ($t\in[0,1]$) as a feature map extracted from the input image, the parameter of the proposed filter $\kappa$ is defined as:

\begin{equation}
    \kappa = (\kappa_{max}-\kappa_{m in})e^{-0.69\times e^{-c(t-t_0)}}+\kappa_{min} \label{eq:NLT}    
\end{equation}
where $\kappa_{min}$ and $\kappa_{max}$ are the minimum and the maximum values that the gain $\kappa$ will take, $c$ is the growth rate of the transformation and $t_0$ is the value of $t$ that produces $\kappa_i=(\kappa_{max}-\kappa_{min})/2$.

\begin{figure}[h]
    \centering
    \includegraphics[clip,trim=0.25cm 0 0 0,width = 0.6\textwidth]{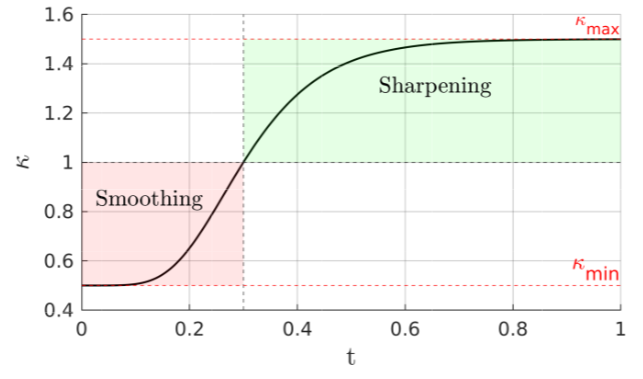}
    \caption{Non-linear transformation, using $\kappa_{min}=0.5, \kappa_{max}=1.5, c=10, t_0 = 0.3$}
    \label{fig:kappa}
\end{figure}

We show an example of the non-linear transformation in \figref{fig:kappa}, where there are two well defined areas in the function: the smoothing region where the $\kappa_{min}\leq\kappa<1$ and the sharpening region where $1<\kappa\leq\kappa_{max}$. We can control the level of smoothing or sharpening by changing the values of $\kappa_{min}$ and $\kappa_{max}$. Setting $\kappa_{min}=1$ will cancel the smoothing effect also setting $\kappa_{max}=1$ will cancel or not produce any sharpening on the image.
In the following subsections we produce content adaptive $\kappa$ by using the non-linear transformation on feature maps.

\subsubsection{Texture guided smoothing and sharpening of face images}

\begin{figure*}[h]
    \centering

\begin{subfigure}{0.31\linewidth}
    \includegraphics[width=\linewidth]{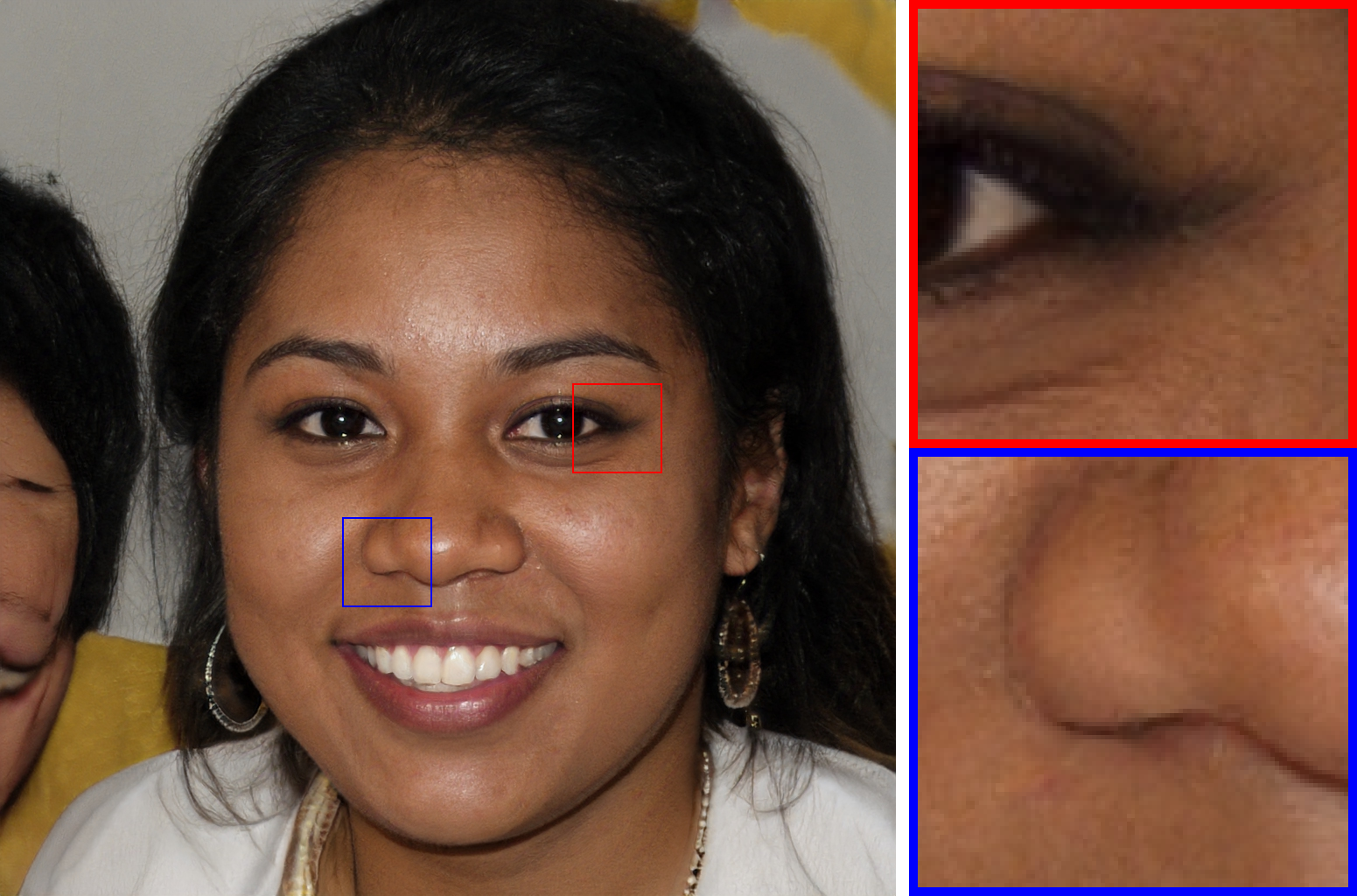}
\end{subfigure}
\begin{subfigure}{0.31\linewidth}
    \includegraphics[width=\linewidth]{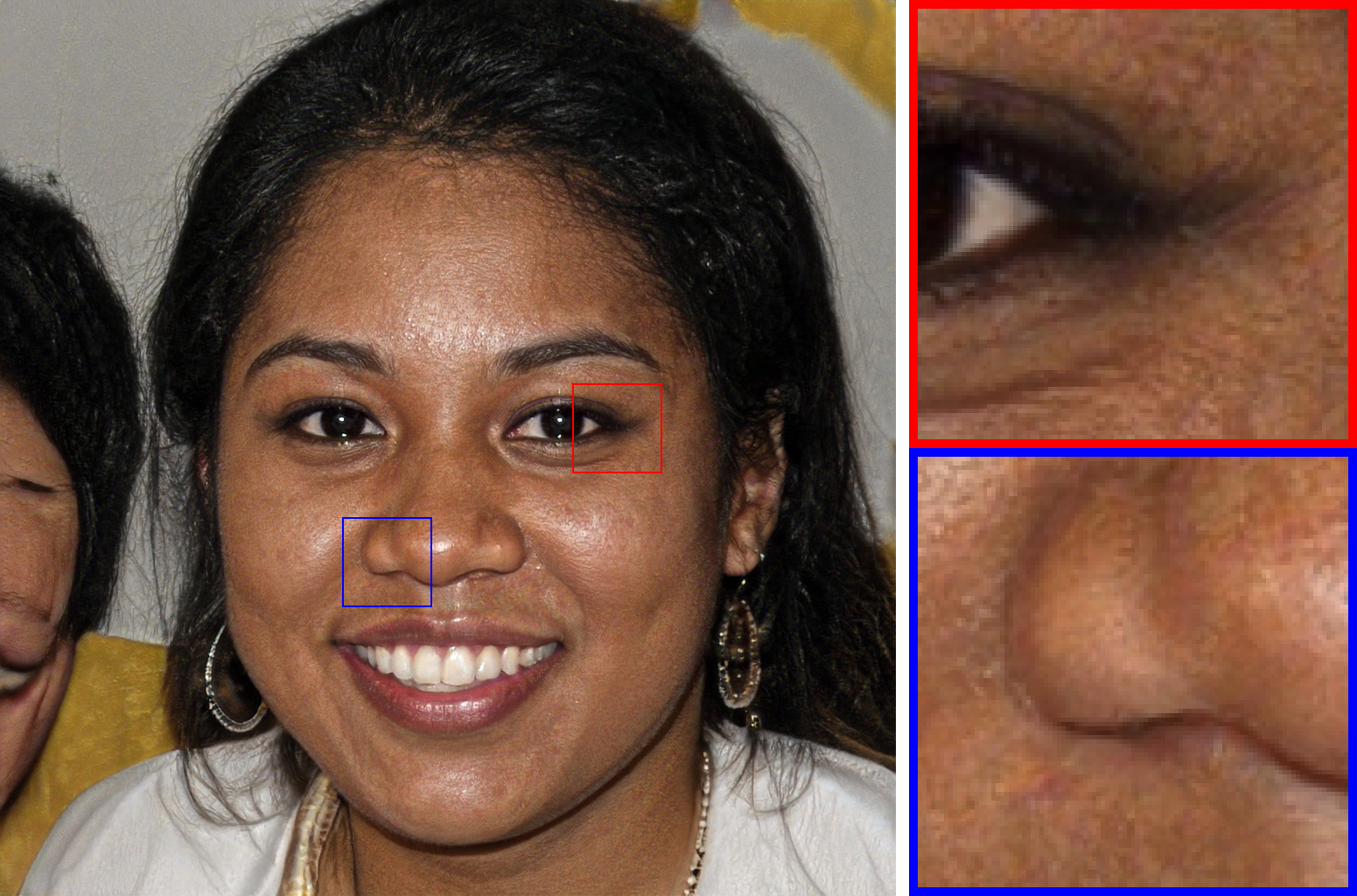}
\end{subfigure}
\begin{subfigure}{0.31\linewidth}
    \includegraphics[width=\linewidth]{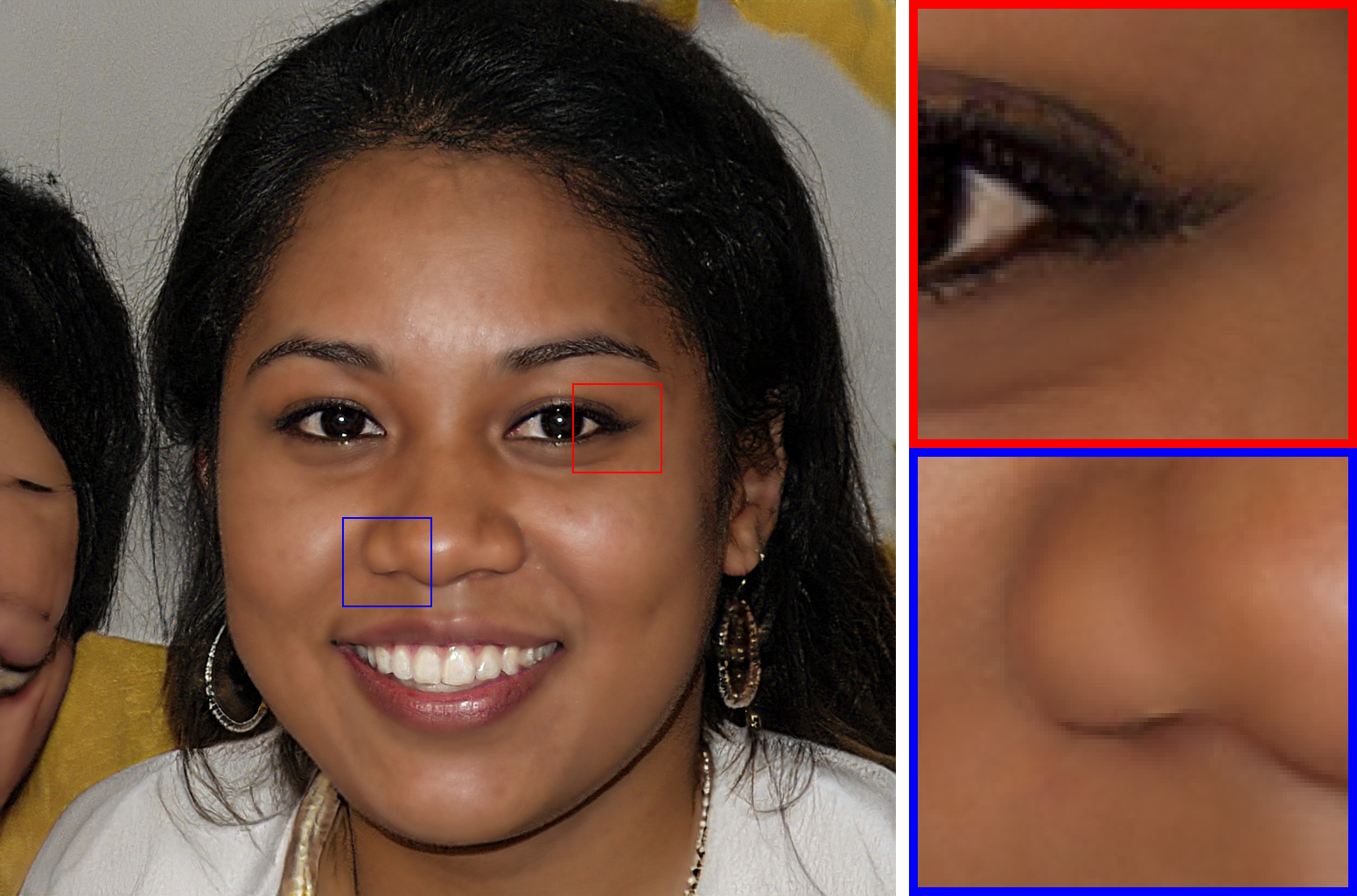}
\end{subfigure}

\begin{subfigure}{0.31\linewidth}
    \includegraphics[width=\linewidth]{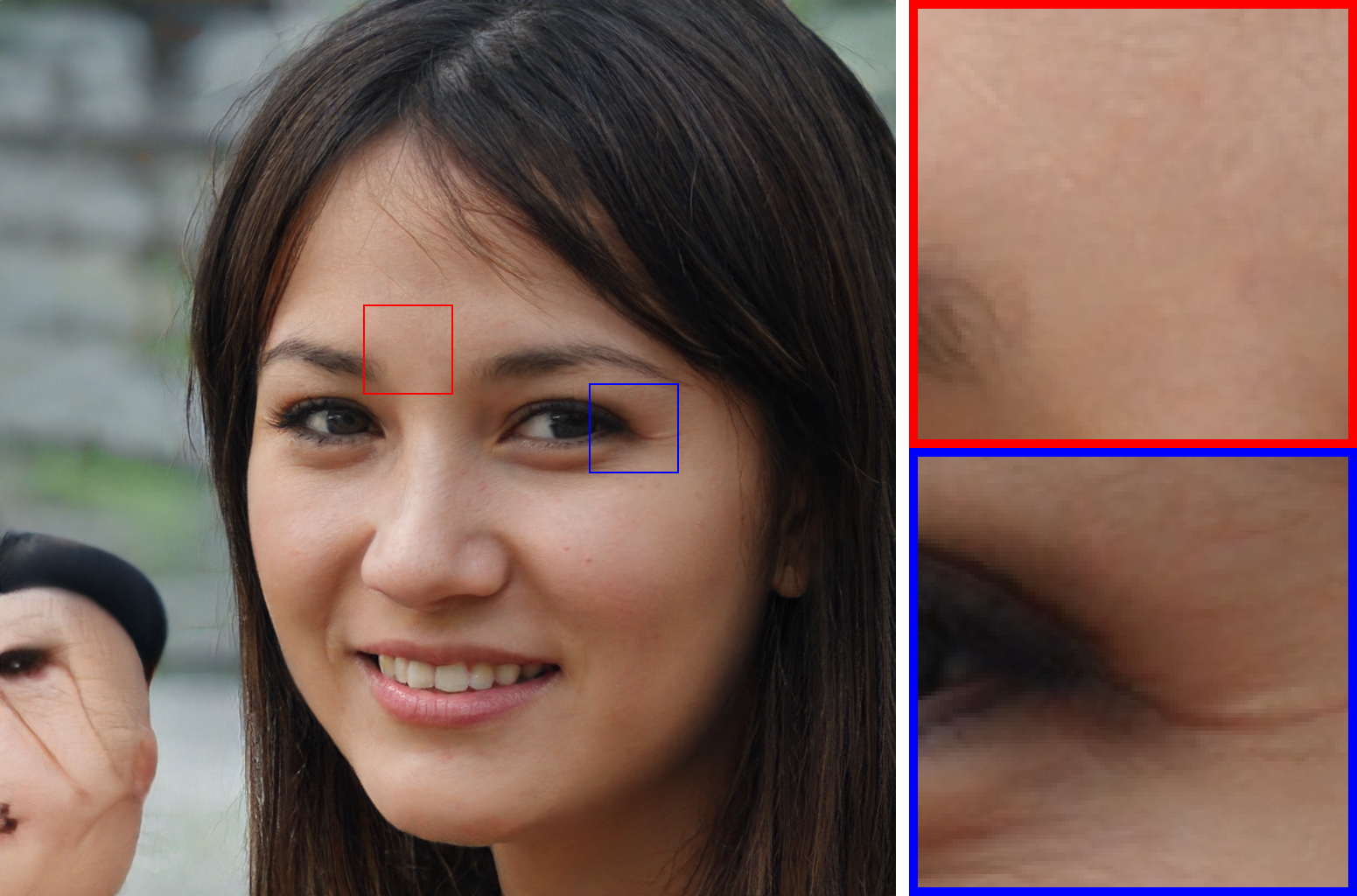}
    \caption{Original image}
\end{subfigure}
\begin{subfigure}{0.31\linewidth}
    \includegraphics[width=\linewidth]{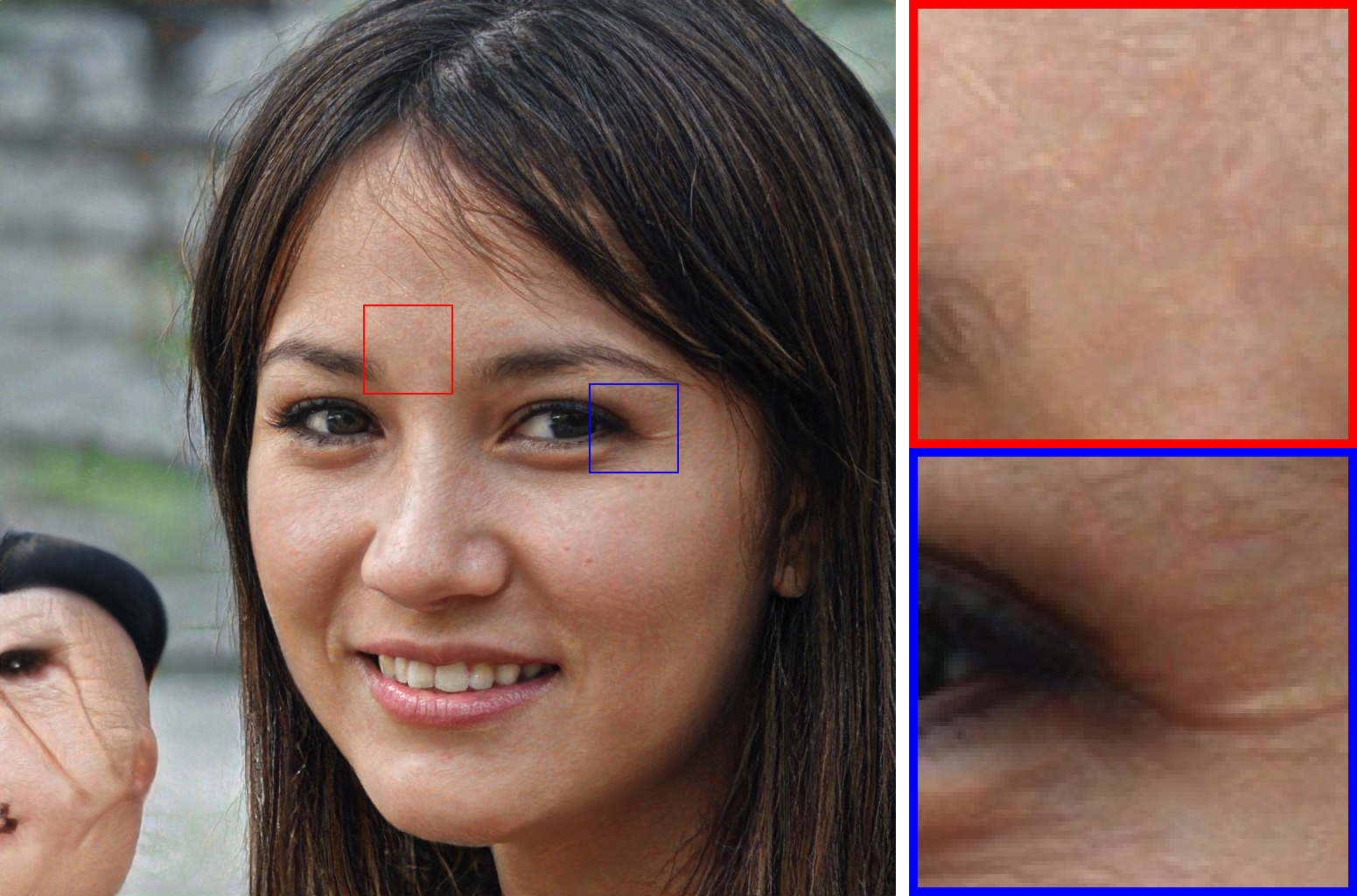}
    \caption{Unsharp masking}
\end{subfigure}
\begin{subfigure}{0.31
\linewidth}
    \includegraphics[width=\linewidth]{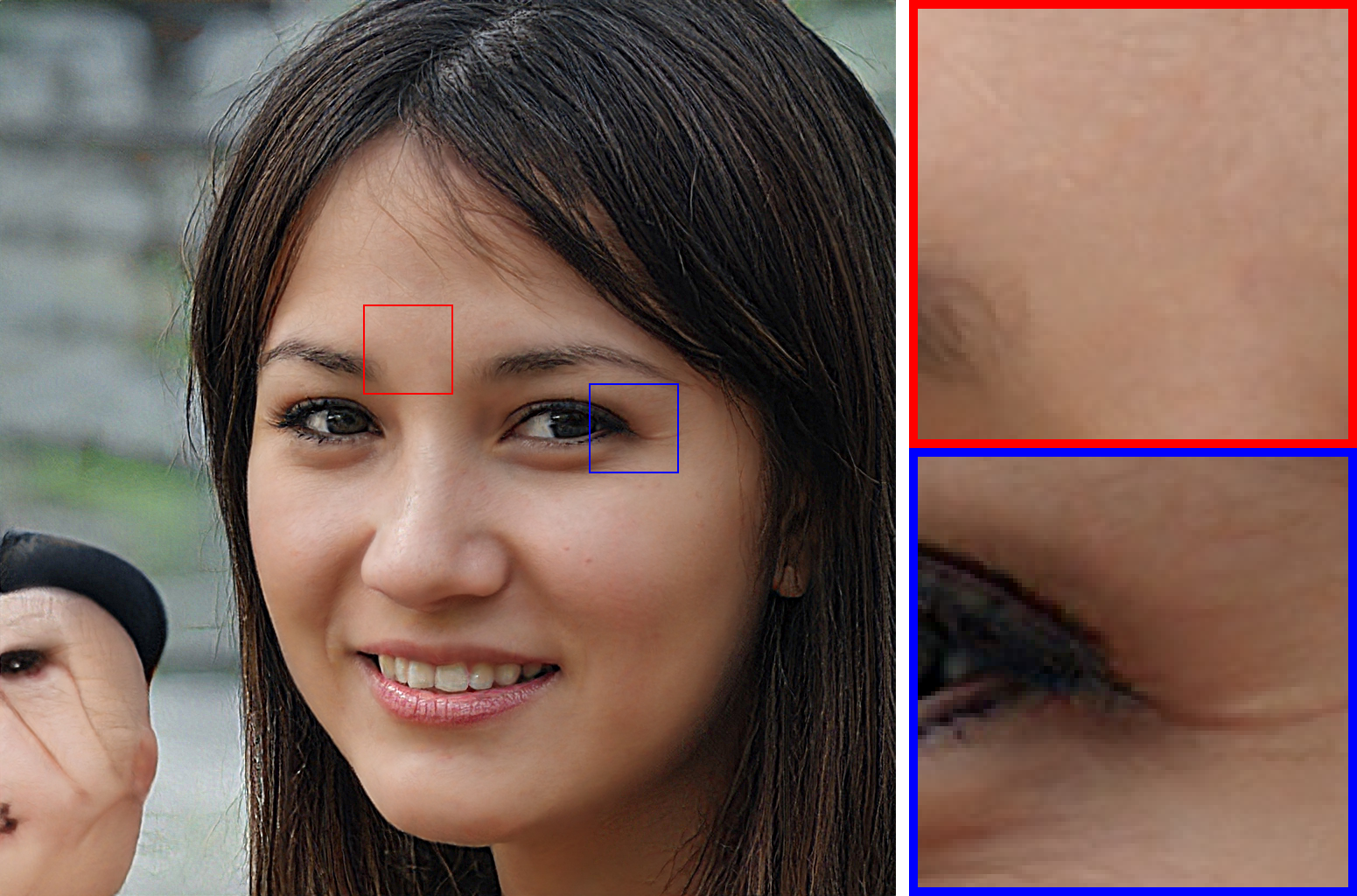}
    \caption{Proposed filter}
\end{subfigure}

\caption{ Face enhancement results. (a) Original image. (b) UM ($\gamma = 1.1$). (c) The proposed filter ($r=3, \epsilon = 0.01, N_{iter}=1, \kappa_{max}=5, \kappa_{min}=0.1$). The proposed method does not sharpen skin regions producing more aesthetically pleasing result.}
\label{fig:faces_compa}
\end{figure*}

\begin{figure*}[h]
\centering
\begin{tikzpicture}[
    zoomboxarray,
    zoomboxarray columns=1,
    zoomboxarray rows=2,
    zoomboxes below,
    zoombox paths/.append style={line width=0.5pt}
]
    \node [image node] { \includegraphics[width=0.19\linewidth]{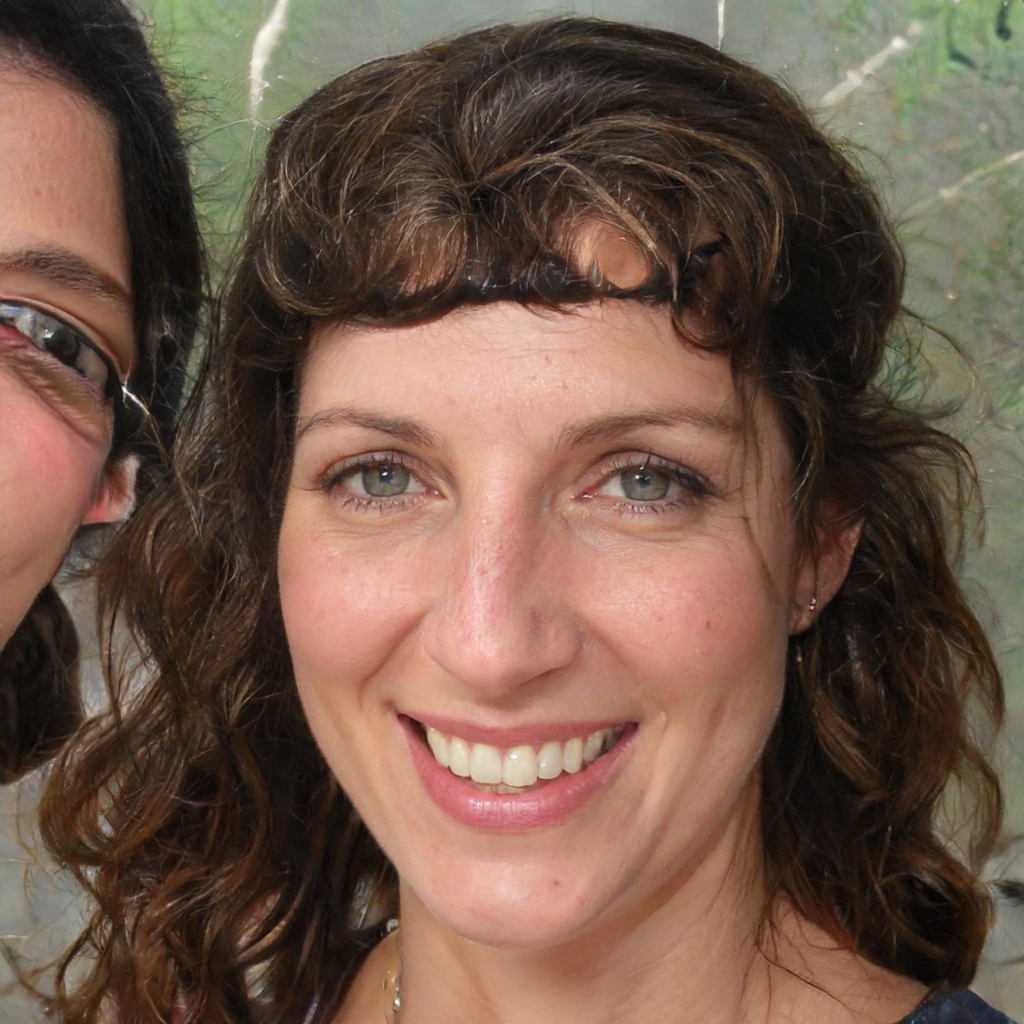} };
    \zoombox[color code = blue,magnification=5]{0.35,0.55}
        \zoombox[color code = green,magnification=5]{0.65,0.4}
\end{tikzpicture}
\begin{tikzpicture}[
    zoomboxarray,
    zoomboxarray columns=1,
    zoomboxarray rows=2,
    zoomboxes below,
    zoombox paths/.append style={line width=0.5pt}
]
    \node [image node] { \includegraphics[width=0.19\linewidth]{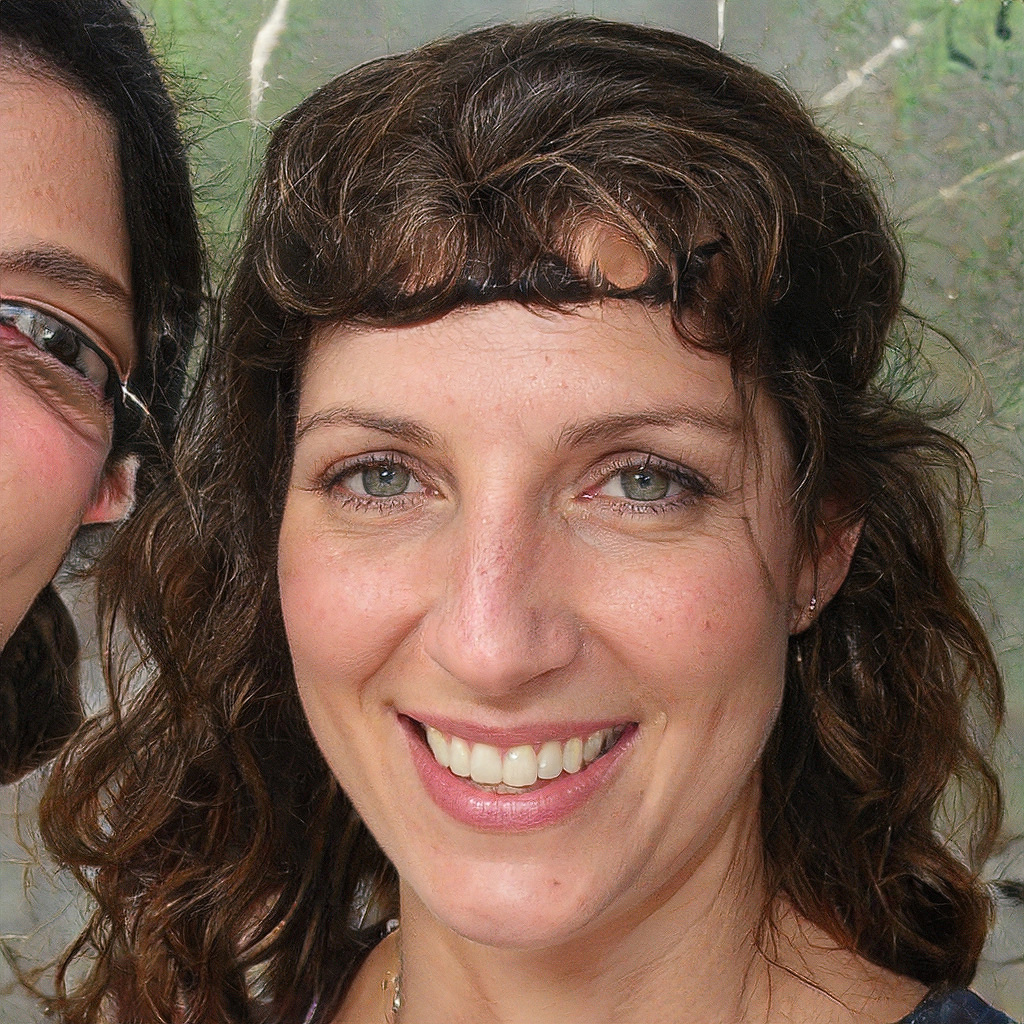} };
    \zoombox[color code = blue,magnification=5]{0.35,0.55}
        \zoombox[color code = green,magnification=5]{0.65,0.4}
\end{tikzpicture}
\begin{tikzpicture}[
    zoomboxarray,
    zoomboxarray columns=1,
    zoomboxarray rows=2,
    zoomboxes below,
    zoombox paths/.append style={line width=0.5pt}
]
    \node [image node] { \includegraphics[width=0.19\linewidth]{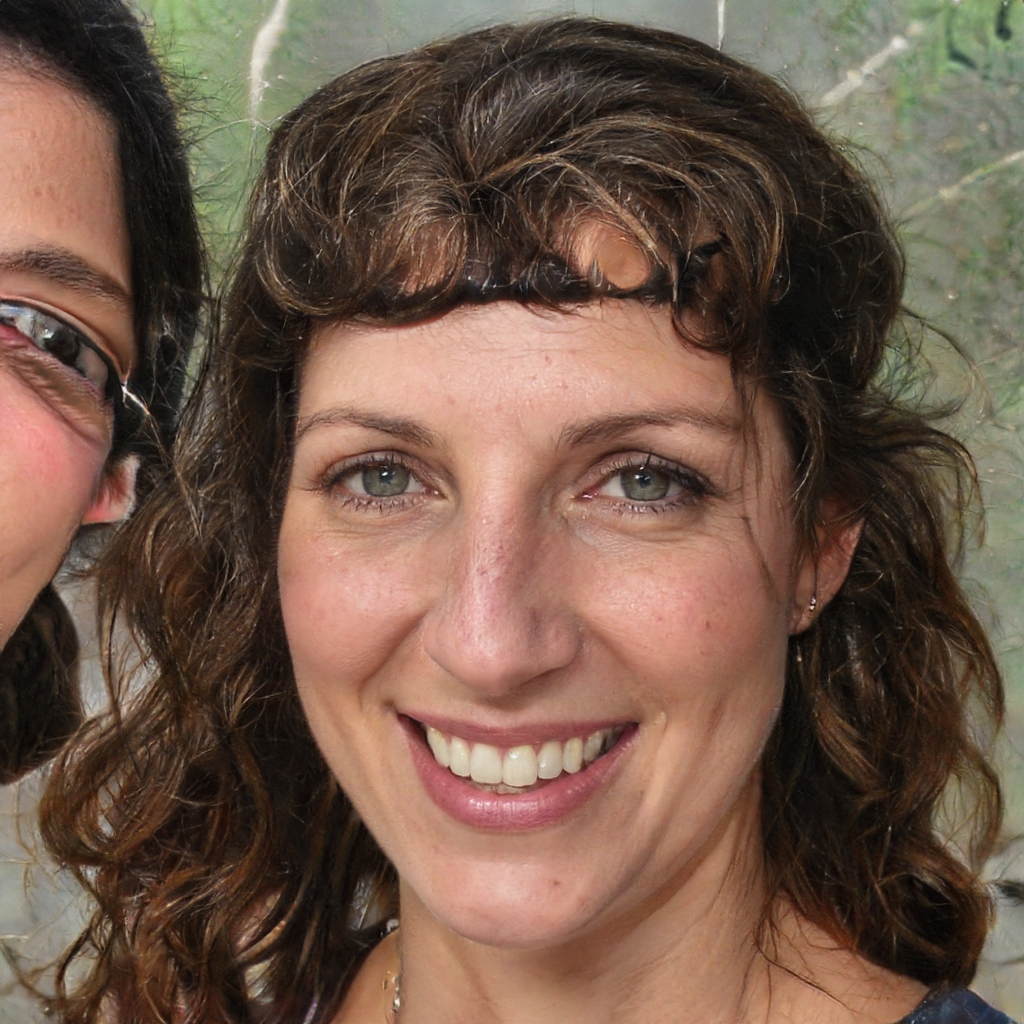} };
    \zoombox[color code = blue,magnification=5]{0.35,0.55}
    \zoombox[color code = green,magnification=5]{0.65,0.4}

\end{tikzpicture}
\begin{tikzpicture}[
    zoomboxarray,
    zoomboxarray columns=1,
    zoomboxarray rows=2,
    zoomboxes below,
    zoombox paths/.append style={line width=0.5pt}
]
    \node [image node] { \includegraphics[width=0.19\linewidth]{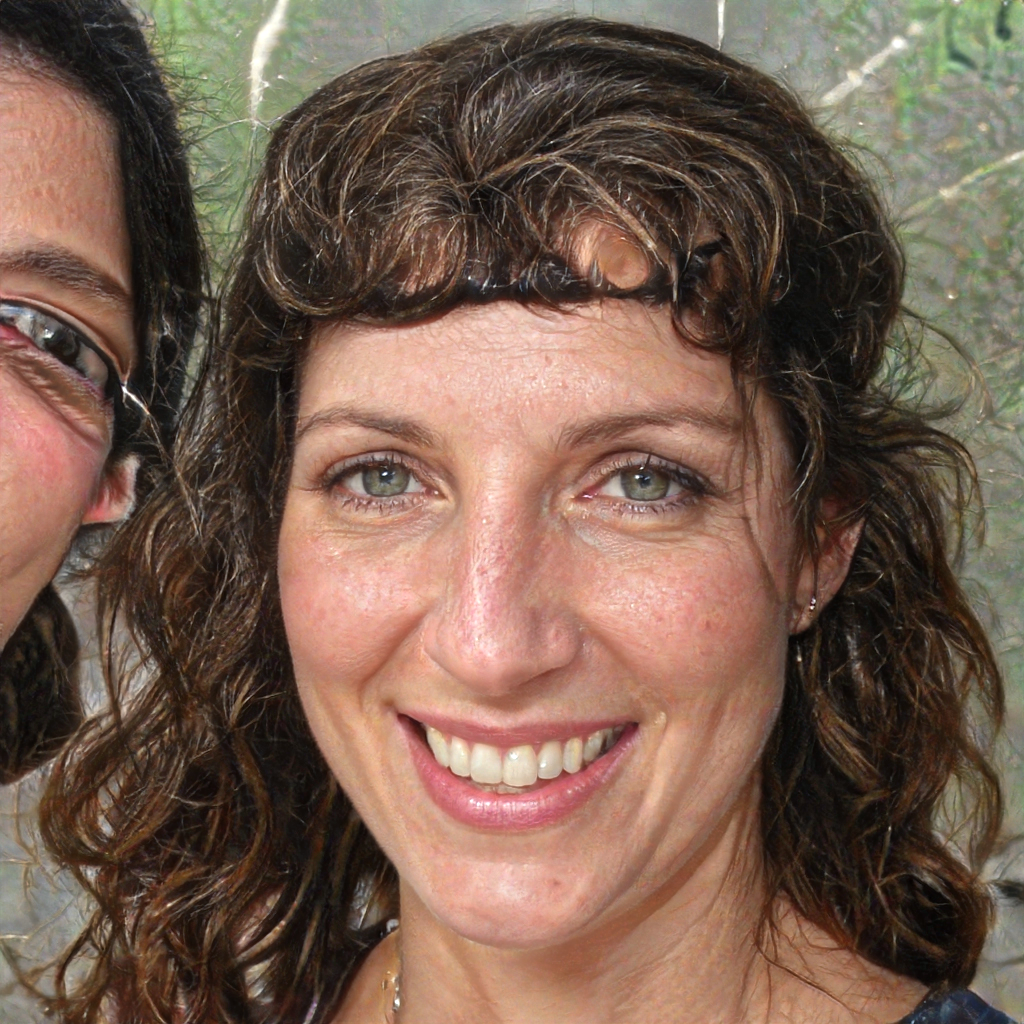} };
    \zoombox[color code = blue,magnification=5]{0.35,0.55}
        \zoombox[color code = green,magnification=5]{0.65,0.4}
\end{tikzpicture}
\begin{tikzpicture}[
    zoomboxarray,
    zoomboxarray columns=1,
    zoomboxarray rows=2,
    zoomboxes below,
    zoombox paths/.append style={line width=0.5pt}
]
    \node [image node] { \includegraphics[width=0.19\linewidth]{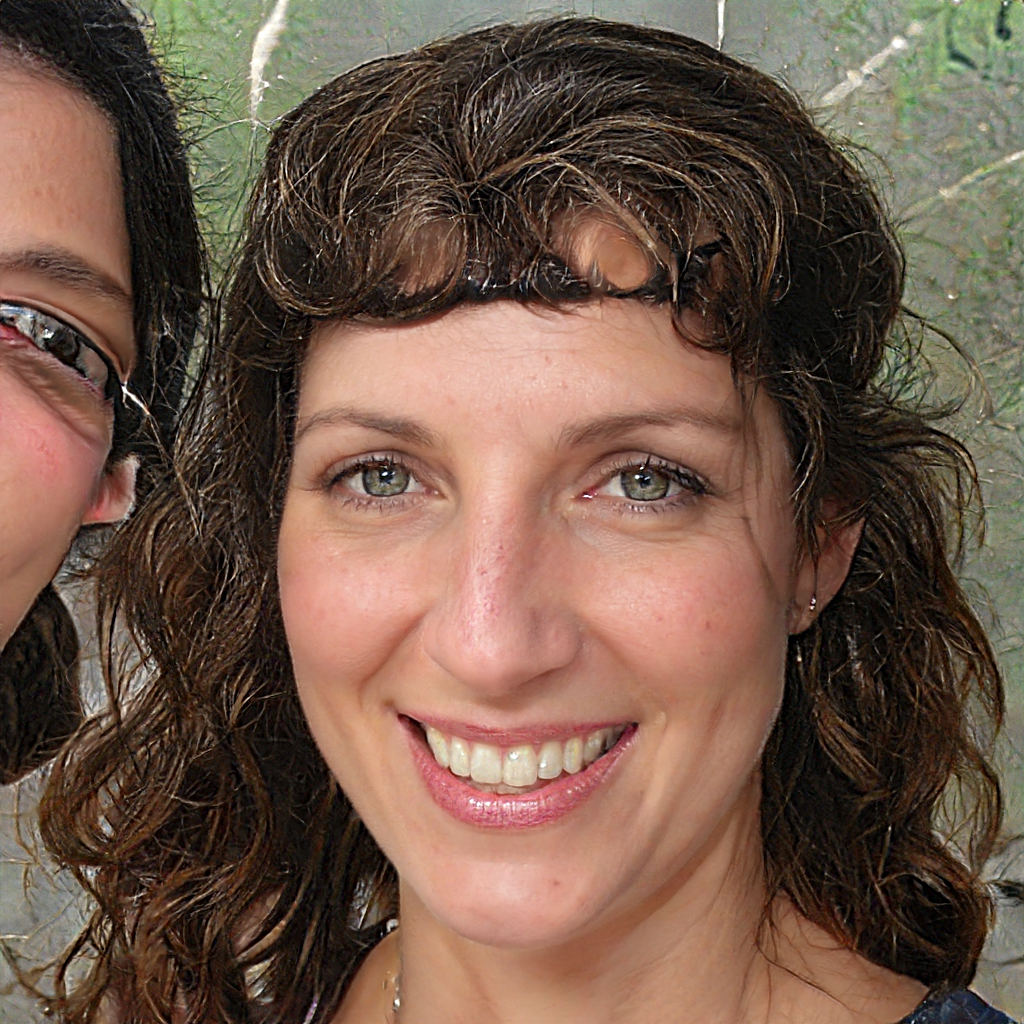} };
    \zoombox[color code = blue,magnification=5]{0.35,0.55}
    \zoombox[color code = green,magnification=5]{0.65,0.4}
\end{tikzpicture}
\caption{Face enhancement. (a, b) Original images. (c, d) CAS results. (e,f) GUM results. (g,h) UM results. (i,j) Proposed filter results.}
\label{fig:comparison_faces_moremethods}
\end{figure*}

A challenge in sharpening portraits by a non-adaptive unsharp masking algorithm is that undesirable effect on skin regions is usually produced. Examples are shown in Figures \ref{fig:faces_compa}(b) and \ref{fig:comparison_faces_moremethods}(g,h) in which the skin part of the image is sharpened. To solve this problem, we first estimate a binary skin map using one of the many algorithms for skin segmentation, e.g., \cite{Dahmani2020, buza2017skin, Phung2003}. We then transform the binary skin map by using the non-linear transformation in \eqref{eq:NLT} to obtain pixel-adaptive $\kappa$, which is used in the proposed filter to sharpen non-skin regions only while gently smoothing the skin region to produce a notable face enhancement.

In \figref{fig:faces_compa} we compare the result of the proposed method against the popular unsharp masking (UM). UM algorithm has a fixed gain ($\gamma$) to amplify the high frequency components of the image. It can be seen that our method produces a more aesthetically pleasing result in skin region while still sharpening the non-skin region. 



In \figref{fig:comparison_faces_moremethods} we compare our method with other state-of-the-art sharpening methods such as generalized unsharp masking (GUM) \cite{deng2010generalized}, unsharp masking \cite{gonzalez2018} and contrast adaptive sharpening (CAS)\footnote{https://www.amd.com/en/technologies/radeon-software-fidelityfx}. Our filter produces a more natural and aesthetically appealing effect on the image than UM and its performance is similar to GUM and CAS.

\begin{figure}[h!]
\centering
\begin{tikzpicture}[
    zoomboxarray,
    zoomboxarray columns=1,
    zoomboxarray rows=2,
    connect zoomboxes,
    zoombox paths/.append style={line width=0.5pt}
]
    \node [image node] { \includegraphics[width=0.24\columnwidth]{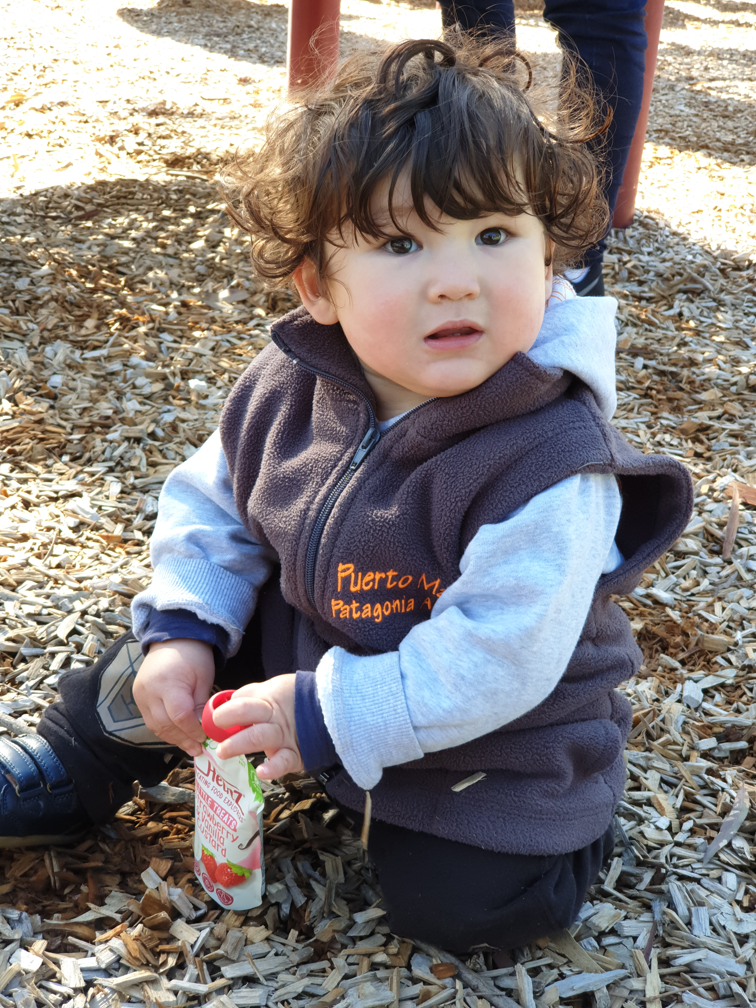} };
    \zoombox[color code = blue,magnification=5]{0.3,0.65}
    \zoombox[color code = green,magnification=5]{0.5,0.4}
\end{tikzpicture}
\begin{tikzpicture}[
    zoomboxarray,
    zoomboxarray columns=1,
    zoomboxarray rows=2,
    connect zoomboxes,
    zoombox paths/.append style={line width=0.5pt}
]
    \node [image node] { \includegraphics[width=0.24\columnwidth]{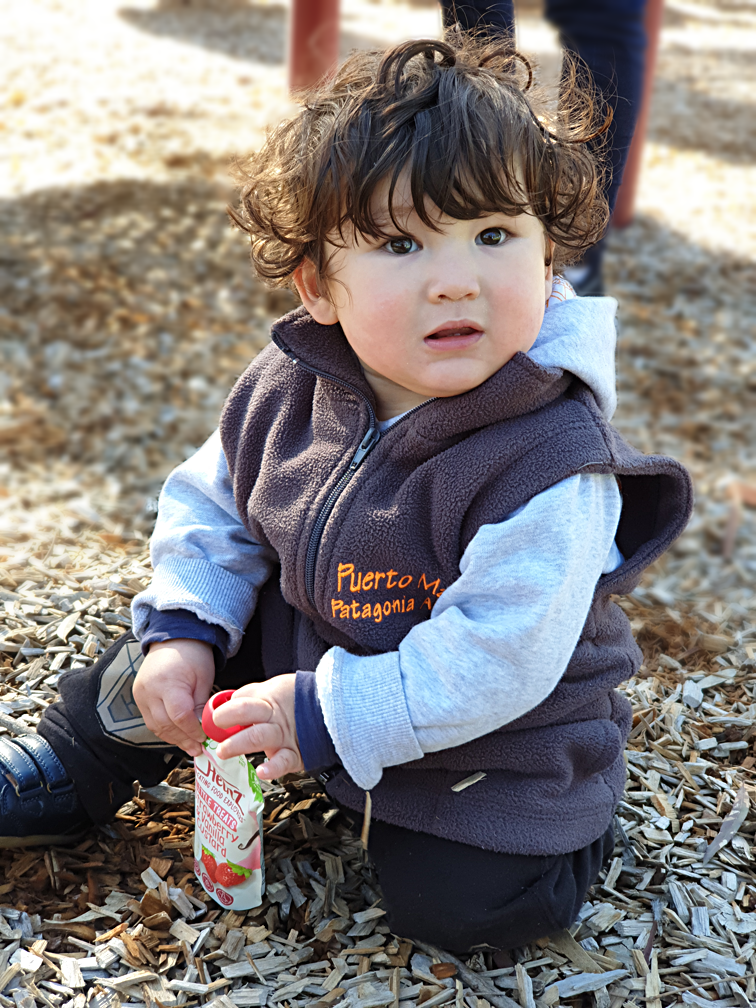} };
    \zoombox[color code = blue,magnification=5]{0.3,0.65}
    \zoombox[color code = green,magnification=5]{0.5,0.4}
\end{tikzpicture}

\caption{Shallow depth of field (SDoF) effect. (a,b) Original image. (c,d) SDoF using the proposed filter ($r=3, \epsilon = 10, N_{iter}=1, \kappa_{max}=2, \kappa_{min}=0$). Our filter not only smooths the background, it also sharpens the object to correct small movements and out of focus. }
\label{fig:robertsdof}
\end{figure}

\subsubsection{Depth guided smoothing for shallow depth of field}
\label{subsec:SDof}
Modern mobile phones can have multiple high resolution cameras to capture high quality images which can be used to estimate the depth map of a scene. We combine this capability and the proposed algorithm to produce a shallow depth of field (SDoF) effect which is frequently used to emphasize the main object. Traditionally, SDoF effect is achieved by using a lens with a large aperture in a SLR camera. The proposed algorithm of pixel-adaptive smoothing and sharpening can be used as post-processing tool to create the SDoF effect. The idea is to obtain the depth map from the phone and perform the non-linear transformation to determine $\kappa$ which is used in the filtering process. In this experiment, we used a Samsung Galaxy Note 9 phone in live focus mode to capture the image and depth map information.  We note that methods such as \cite{liu2015deep, godard2017unsupervised} can be used to estimate the depth map from a single image if only one camera is available.

The depth map $D$ is a gray-scale image with values in the range [0,1]. A closer object in the scene has a smaller D value. So the feature map is defined as $t=1-D$. Applying the non-linear transformation, the depth information is mapped to $\kappa$ which is a decreasing function of $D$ leading to progressively smoothing effect as the distance increases. At the same time the closer objects are sharpened to correct slight out of focus or blur. 

\begin{figure}[h!]
\centering
\begin{tikzpicture}[
    zoomboxarray,
    zoomboxarray columns=1,
    zoomboxarray rows=2,
    connect zoomboxes,
    zoombox paths/.append style={line width=0.5pt}
]
    \node [image node] { \includegraphics[width=0.24\columnwidth]{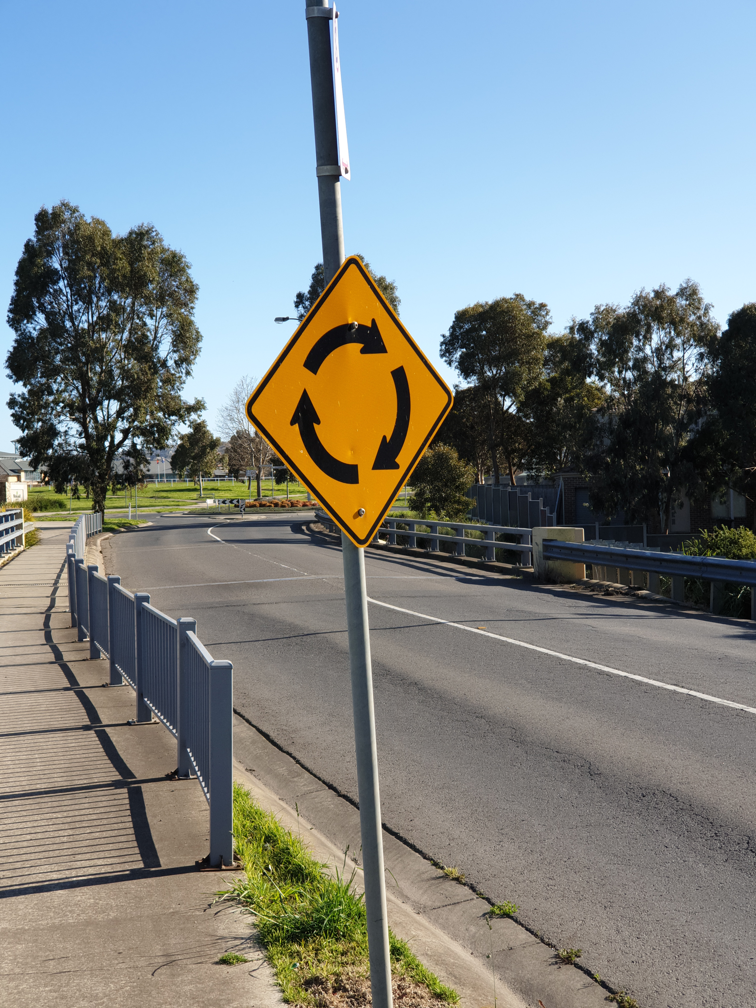} };
    \zoombox[color code = blue,magnification=5]{0.3,0.65}
        \zoombox[color code = green,magnification=5]{0.35,0.3}
\end{tikzpicture}
\begin{tikzpicture}[
    zoomboxarray,
    zoomboxarray columns=1,
    zoomboxarray rows=2,
    connect zoomboxes,
    zoombox paths/.append style={line width=0.5pt}
]
    \node [image node] { \includegraphics[width=0.24\columnwidth]{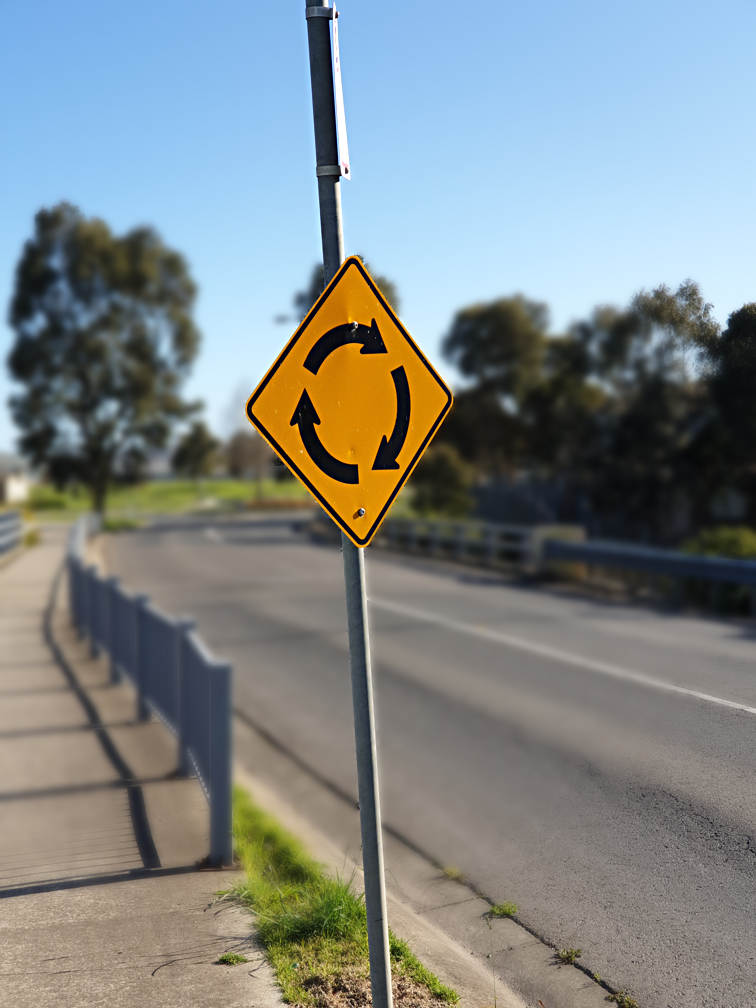} };
    \zoombox[color code = blue,magnification=5]{0.3,0.65}
        \zoombox[color code = green,magnification=5]{0.35,0.3}
        \figurename{jj}
\end{tikzpicture}
\caption{Shallow depth of field (SDoF). (a,b) Original image. (c, d) SDof using the proposed filter ($r=1, \epsilon = 100, N_{iter}=10, \kappa_{max}=2, \kappa_{min}=0$). Our filter not only smooths the background, it also sharpens the object.}
\label{fig:signsdof}
\end{figure}

In \figref{fig:robertsdof} and \figref{fig:signsdof} we show two different results where the foreground is sharpened and the background is smoothed to produce the SDoF effect. We can see that the background is smoothed with a natural appearance to simulate the defocus blur, also the foreground is sharpened reducing the blur due to movement in \figref{fig:robertsdof} and out of focus in \figref{fig:signsdof}. 

\subsubsection{Content-aware seam carving}
\begin{figure*}[h!]
\centering
\begin{subfigure}[b]{0.25\textwidth}
    	\includegraphics[width=\textwidth]{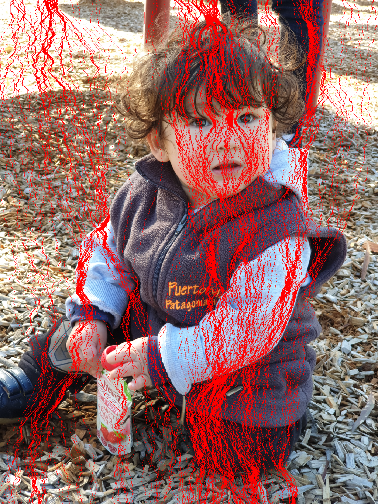}
    	\caption{}
    \end{subfigure}
    \begin{subfigure}[b]{0.213\textwidth}
    	\includegraphics[width=\textwidth]{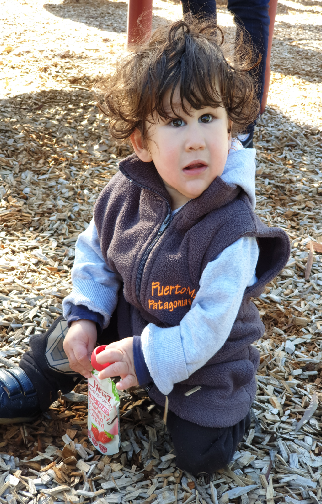}
    	\caption{}
    	\label{fig:SeamCarving_b}
    \end{subfigure}
    \begin{subfigure}[b]{0.25\textwidth}
    	\includegraphics[width=\textwidth]{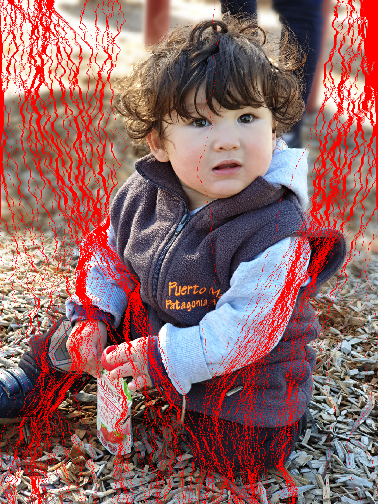}
    	\caption{}
    \end{subfigure}
    \begin{subfigure}[b]{0.213\textwidth}
    	\includegraphics[width=\textwidth]{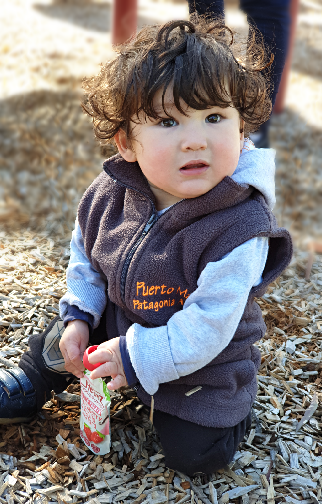}
    	\caption{}
    	\label{fig:SeamCarving_d}
    \end{subfigure} 
	\caption{Seam carving results. (a) Original image and seams to be removed. (b) Removed seams in (a). (c) Image after background smoothing and seam to be removed. (d) Removed seams in (c). The proposed background filter leads to a better result.}
	\label{fig:SeamCarving_ddd}
\end{figure*}

We present an application of the background smoothing algorithm detailed in previous section as a pre-processing step for seam carving. Seam carving was introduced in \cite{Avidan2007} as an effective tool for resizing an image without significant change to main objects. The idea is to delete pixels of unimportant details in an image. The importance of a pixel is measured by a function of gradient. A natural image often contains details such as trees, sand, grass which are usually less important compared to the object of interest such as human. However, a direct application of gradient-based seam-carving may lead to unsatisfactory results. An example is shown in \figref{fig:SeamCarving_b}. A solution to this problem is to use content-aware image resizing. To avoid elimination of information in foreground, we pre-process the input image by using the proposed SDoF algorithm which not only smooths out details in the background but also emphasizes the object of interest by sharpening it. Results are shown in \figref{fig:SeamCarving_d} which shows that after SDoF filtering the seams are not running over the boy's face. As a result, the seam-carving algorithm produces a better output image. 

\subsubsection{Blurriness guided sharpening and smoothing}

When an object in a scene is outside the focal plane it is defocused with a blur level directly proportional to the distance from the focal plane \cite{pentland1987new}. Due to the limited depth of field intrinsic in most optical systems, the defocus blur is present in most of images. When the depth information is not available, the defocus blur becomes the simplest depth cue in an image \cite{mather1996image} and it is widely used by photographers to make the main object of the scene to stand out from the background. Sometimes, due to wrong focal settings, images need to be sharpened or deblured to achieve a more pleasant result. 

When an image has defocus blur present, the sharpening process becomes a challenge. Sharpening highly defocused regions produces artifacts and sharpening in focus regions can lead to over-sharpening. To tackle these problems we propose a method  by sharpening and smoothing an image adaptively using the defocus-blur map to compute a spatially varying $\kappa$ map. The key behind the success of our method is the estimation of the defocus-blur map. 

Methods have been developed to estimate the defocus-blur map. Some of the methods use multiple images or special hardware \cite{Levin2007, Vu2014, Zhou2009} while others use a single image. These methods fall into two main categories. The first one is the traditional image processing methods such as \cite{Chakrabarti2010, Zhu2013a} which perform a frequency domain analysis to estimate the defocus-blur map or \cite{Zhang2016, Hu2006} which use changes in gradient to estimate the blur level at edges and then interpolate those level to the rest of the image using a matting algorithm. The second one is the machine learning based methods such as \cite{Lee2019} which uses an end-to-end CNN to estimate the defocus map and \cite{DAndres2016} which estimates the blur map by using a regression tree field (RTF) model. 

In this work we use entropy, which is a measure of pixel variation in a local area, as an indicator for the defocus. The defocus-blur map $D$ at a pixel location is defined as the entropy of a patch centered at that particular pixel. The map is then refined by using the guided filter \cite{he2012guided} which uses the original image as the guidance. The refined defocus-blur map is non-linearly transformed by equation \eqref{eq:gompertz} to obtain the desired $\kappa$ map. The proposed filter with the pixel-adaptive $\kappa$ is applied to the image. The result is adaptive smoothing-sharpening based on the local entropy information as a measure of defocus-blur.

\begin{figure*}[h!]
\centering
\begin{subfigure}{0.31\linewidth}
    	\includegraphics[width=\linewidth]{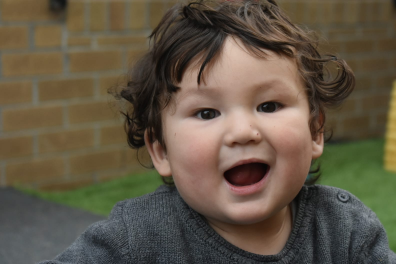}
    	\caption{Original image}
    \end{subfigure}
    \begin{subfigure}{0.31\linewidth}
    	\includegraphics[width=\linewidth]{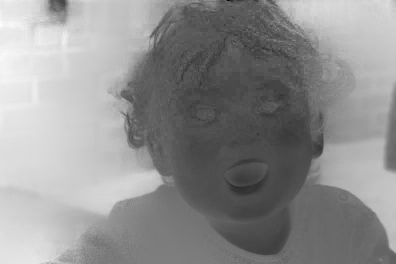}
    	\caption{Entropy map}
    	\label{fig:Blur_guided_b}
    \end{subfigure}
    \begin{subfigure}{0.31\linewidth}
    	\includegraphics[width=\linewidth]{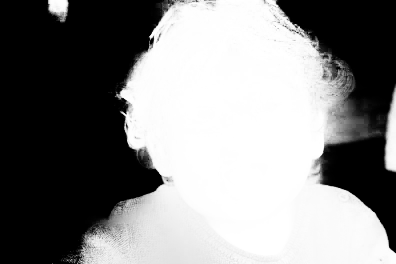}
    	\caption{$\kappa$ map}
    	\label{fig:Blur_guided_c}
    \end{subfigure}
    \begin{subfigure}{0.31\linewidth}
    	\includegraphics[width=\linewidth]{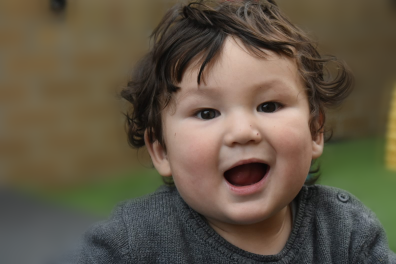}
    	\caption{Blurring defocused regions}
    	\label{fig:Blur_guided_d}
   \end{subfigure}
    \begin{subfigure}{0.31\linewidth}
    	\includegraphics[width=\linewidth]{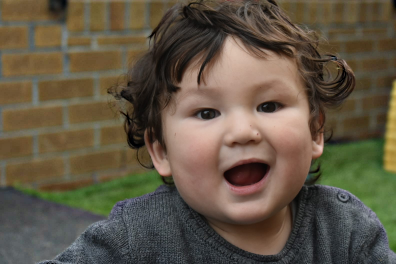}
    	\caption{Sharpening defocused regions}
    \end{subfigure}
    \begin{subfigure}{0.31\linewidth}
    	\includegraphics[width=\linewidth]{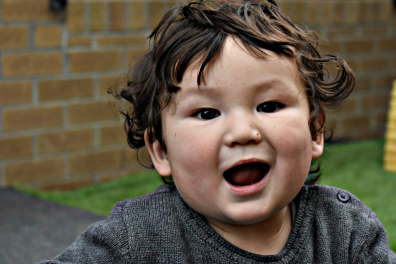}
    	\caption{Sharpening with a fixed $\kappa$}
    \end{subfigure} 
	\caption{Sharpening/smoothing guided by defocus. (a) Original image. (b) Refined entropy map. (c) $\kappa$ map. (d) Result of blurring defocused regions to produce a shallow depth of field. (e) Sharpening defocused regions to increase the depth of field. (f) Sharpening the whole image with a fixed $\kappa$.}
	\label{fig:Blur_guided}
\end{figure*}

In \figref{fig:Blur_guided} we first show an image which contains different levels of focal blur. The boy's face (image (a)) is in focus while the background is slightly out of focus. The refined entropy map (image (b)) is produced by the MATLAB function \texttt{entropyfilt} with a window size of 33x33 pixels and the result is refined by using guided filter \cite{he2012guided} ($ r = 32, \epsilon = 0.01$). The $\kappa$ map is shown in image (c). 

In the second row of \figref{fig:Blur_guided} we show 3 different results. Image (d) is the result of smoothing areas with some degrees of focal blur. The background is successfully blurred while the boy's face remains sharp leading to a pleasing SDoF effect. Image (e) is the result of sharpening the out-of-focus regions to increase the depth of field. The background appears sharper producing the sensation that the scene's depth is slightly larger than the original. In both cases, the boy's face which is in focus, is not changed (achieved by setting $\kappa=1$). However, we should point out that the proposed filter can be easily configured to sharpen the in-focus objects by adjusting the parameters of the non-linear transformation. As a comparison, image (f) shows the effect of sharpening with a fixed $\kappa$ value for the whole image, leading to an image of non natural appearance. 

\begin{figure}[h!]
\centering
\begin{subfigure}{0.4\columnwidth}
    	\includegraphics[width=\linewidth]{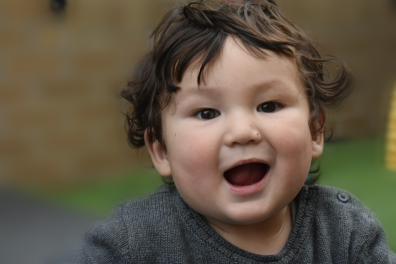}
    	\caption{Proposed method}
    \end{subfigure}
    \begin{subfigure}{0.4\columnwidth}
    	\includegraphics[width=\linewidth]{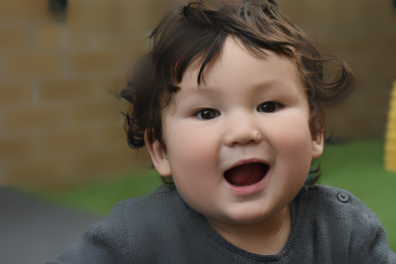}
    	\caption{GDGF \cite{gradientDomainGF}}
    \end{subfigure}
    \begin{subfigure}{0.4\columnwidth}
    	\includegraphics[width=\linewidth]{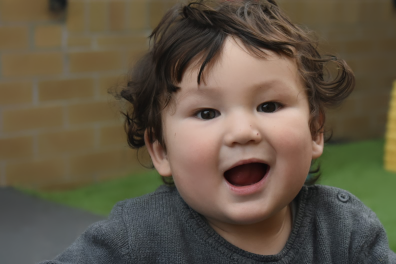}
    	\caption{WGF \cite{li2015weighted}}
    \end{subfigure}
    \begin{subfigure}{0.4\columnwidth}
    	\includegraphics[width=\linewidth]{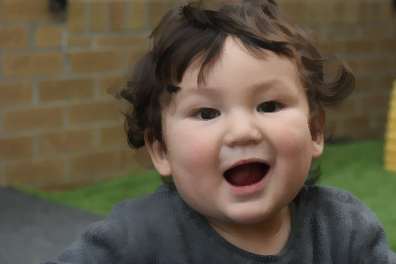}
    	\caption{SWGF \cite{sideWindowFitering}}
   \end{subfigure}

	\caption{Background smoothing. Comparison with other guided filters (a) Proposed method (same as Fig.13(d)). (b) GDGF \cite{gradientDomainGF} ($ r = 7, \epsilon=0.01$). (c) WGF \cite{li2015weighted} ($ r = 7, \epsilon=0.001$) (d) SWGF \cite{sideWindowFitering} ($r = 3, N_{iter}=3$).}
	\label{fig:guided_filter_comparisons}
\end{figure}

To further demonstrate the advantage of the proposed smoothing-sharpening filter, we compare results from the gradient domain guided filter (GDGF) \cite{gradientDomainGF}, the weighted guided filter (WGF) \cite{li2015weighted}, and the side window guided filter (SWGF\footnote{\href{https://github.com/YuanhaoGong/SideWindowFilter}{https://github.com/YuanhaoGong/SideWindowFilter}}) \cite{sideWindowFitering}. We tune parameters of these 3 filters such that the background is maximally smoothed while preserving information of the face shown in \figref{fig:Blur_guided}(a) as much as possible. Results are shown in \figref{fig:guided_filter_comparisons} which clearly shows that the proposed filter has the best capability for blurring the background while preserving the information of the face which is almost unchanged. On the other hand, both GDGF and WGF can blur the background to some extend at the cost of smoothing the face. For the SWGF, we have to choose a relatively small patch size to prevent the face being overly smoothed. As a result, there is little effect on smoothing the background.

\subsection{External guided smoothing and sharpening}\label{sec:ext guided filt}

\subsubsection{Flash/no-flash sharpening}

\begin{figure}
    \centering
\begin{subfigure}{0.3\columnwidth}
    \includegraphics[width=\textwidth]{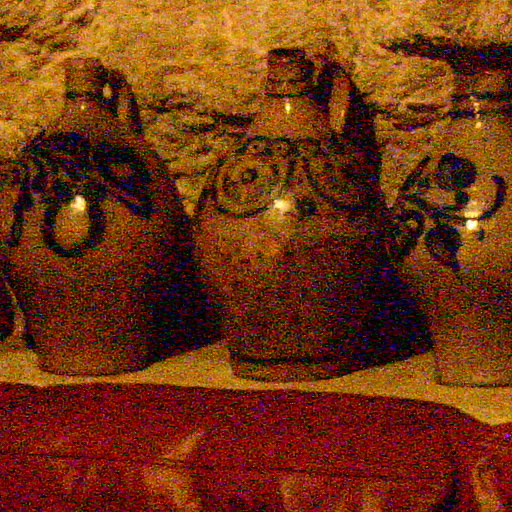}
    \subcaption{Without-flash}
    \label{fig:flash_no_flash_a}
\end{subfigure}
\begin{subfigure}{0.3\columnwidth}
    \includegraphics[width=\textwidth]{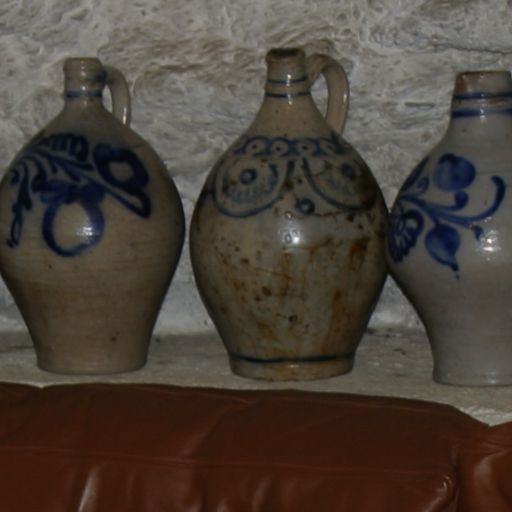}
    \subcaption{With-flash}
\end{subfigure}
\begin{subfigure}{0.3\columnwidth}
    \includegraphics[width=\textwidth]{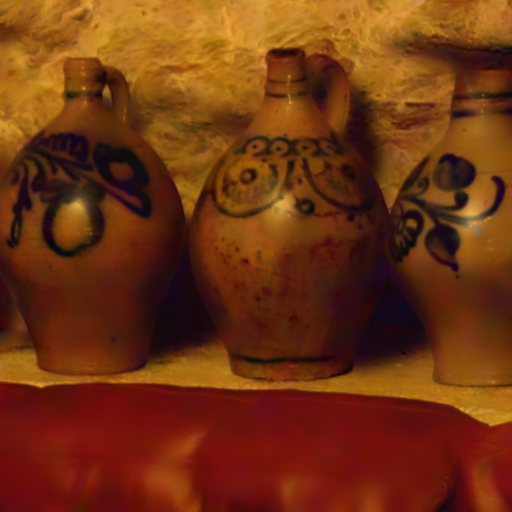}
    \subcaption{Original GF result}
\end{subfigure}
\begin{subfigure}{0.3\columnwidth}
    \includegraphics[width=\textwidth]{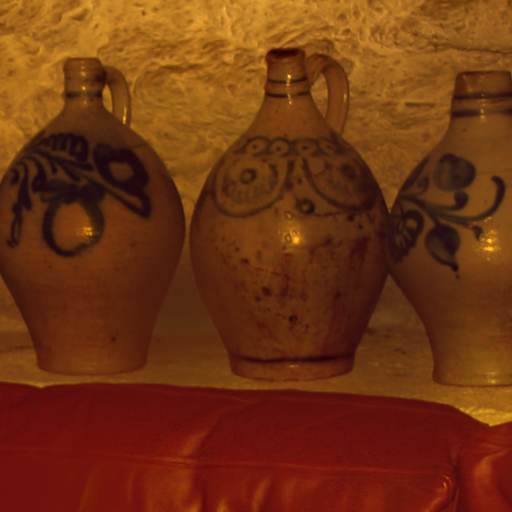}
    \subcaption{Iterative GF result}
\end{subfigure}
\begin{subfigure}{0.3\columnwidth}
    \includegraphics[width=\textwidth]{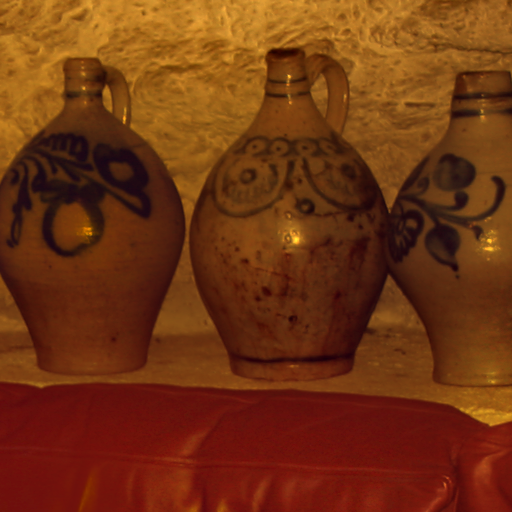}
    \subcaption{Proposed result $\kappa=10$}
\end{subfigure}
\begin{subfigure}{0.3\columnwidth}
    \includegraphics[width=\textwidth]{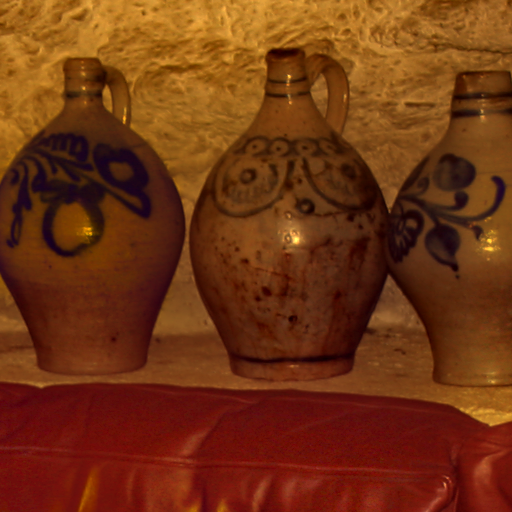}
    \subcaption{Proposed result $\kappa=100$}
\end{subfigure}
\caption{Flash/no flash sharpening. (a) Image without flash. (b) Image with flash. (c) GF ($r=8, \epsilon = 0.004$). (d) Iterative GF ($r=25, \epsilon = 10^{-6}, N_{iter}=10$). (e) The proposed filter ($r=25, \epsilon = 10^{-6}, N_{iter}=10, s=1, \kappa=10 $). (f) The proposed filter ($r=25, \epsilon = 10^{-6}, N_{iter}=10, s=1, \kappa=100 $).}
\label{fig:flash_no_flash}
\end{figure}

In this section we demonstrate another application of the proposed filter where guided smoothing and sharpening is required. For example,  a picture taken under low light condition contains a high level of noise due to the use of high ISO setting. One of the successful applications of the original guided filter \cite{he2012guided} is
in processing images captured under low light condition without using the flash. One such image is shown in \figref{fig:flash_no_flash_a}. The idea is to use another image captured with flash-on as a guidance to enhance the one without flash. In the original implementation, the guided filter with parameters $r=8$ and $\epsilon=0.004$ are used. Result is shown in \figref{fig:flash_no_flash}(c), where we can see that noise has been greatly reduced while the color information is preserved. This is however at the cost of loss of details, e.g., details on the wall and on the vases. This is evident when we compare the result with the guidance image (with-flash).

To tackle this problem, we first apply the guided filter in an iterative manner. We represent the filter operation as $J=GF(I,G)$ where $GF$ denotes the guided filter. The iteration is performed as: $J^{(0)}=I$ and $J^{(n)}=GF(J^{(n-1)},G)$. Using the parameter settings $r=25$ and $\epsilon=10^{-6}$, we perform 10 iterations and show the result in \figref{fig:flash_no_flash}(d). Comparing the original GF result with the iterative GF result, we can see that the latter has retained more details of the original scene than the former.

Next, we test the proposed guided smoothing-sharpening filter in the same iterative way, i.e., same parameter settings with 10 iterations. The proposed filter has two extra parameters: $\kappa$ and the scale $s$. For simplicity, we set $s=1$ and $\kappa=10$ and $100$ to study the sharpening effect. Results in \figref{fig:flash_no_flash}(e) and (f)
which show that the proposed filter does indeed produce sharper results than the iterative GF. To make a quantitative comparison, we calculate the total variation of the image. The total variation for image $I$ is defined as 

\begin{equation}
    TV(I)=\sum_{c\in\{R,G,B\}}\sum_{n}|I_{h}^{(c)}(n)|+|I_{v}^{(c)}(n)|
\end{equation}
where $I_{h}^{(c)}(n)$ / $I_{v}^{(c)}(n)$ is the first derivative
of the image along the horizontal/vertical direction at location $n$
and the superscript $c$ is used to indicate the color channel. Since
the total variation is the sum of absolute values of the first derivative,
we can use it as an indication of the sharpness of the image in this
application. The total variations for different settings of $\kappa$
are shown in Table \ref{tab:TV}, where $\kappa=0$ corresponds to the iterative
guided filter. We can see that the sharpness of the image is indeed an
increasing function of $\kappa$.

\begin{table}
\begin{centering}
\caption{The total variation of the image produced by the proposed iterative smoothing-sharpening filter. It is an increasing function of $\kappa$ indicating the image appear to be sharper by using a larger setting of $\kappa$.}
\centering

\renewcommand{\arraystretch}{1}
\label{tab:TV}
\begin{tabular}{c c c c c c}
\\
\toprule
$\kappa$ & 0 & 10 & 50 & 100 & 200\\
\midrule
$TV(\times10^{4})$ & 1.13 & 1.33 & 1.56 & 1.77 & 2.09\\
\bottomrule
\end{tabular}
\par\end{centering}
\end{table}
\renewcommand{\arraystretch}{1}

Another issue is related to the number of iterations. In general,
for $\kappa>1$, more iterations tend to produce a higher degree of
sharpening effect. How to set the number of iterations is application
dependent and it can be a user specified parameter. For the image
shown in \figref{fig:flash_no_flash}, we empirically found that between 5 to 10 iterations and a setting
of $10<\kappa<100$ result in visually pleasing images.

\subsubsection{Pan-sharpening}

 Multi spectral (MS) images usually have low spatial resolution but are rich in spectral information. On the other hand panchromatic images (called PAN images) have low spectral resolution but have  high spatial resolution. For example, the IKONOS and QuickBird imaging sensors capture a PAN image with a spatial resolution of 1 and 0.6 m respectively and a MS image with a spatial resolution of 4 and 2.6 m respectively \cite{Tu2005}.  Pan-sharpening is a technique that combines information of MS images with PAN images to produce a high spatial resolution image with large spectral information. Pan-sharpening is a useful tool in many remote sensing applications. 
\begin{figure*}
    \centering
\begin{subfigure}{0.3\linewidth}
    \includegraphics[width=\linewidth]{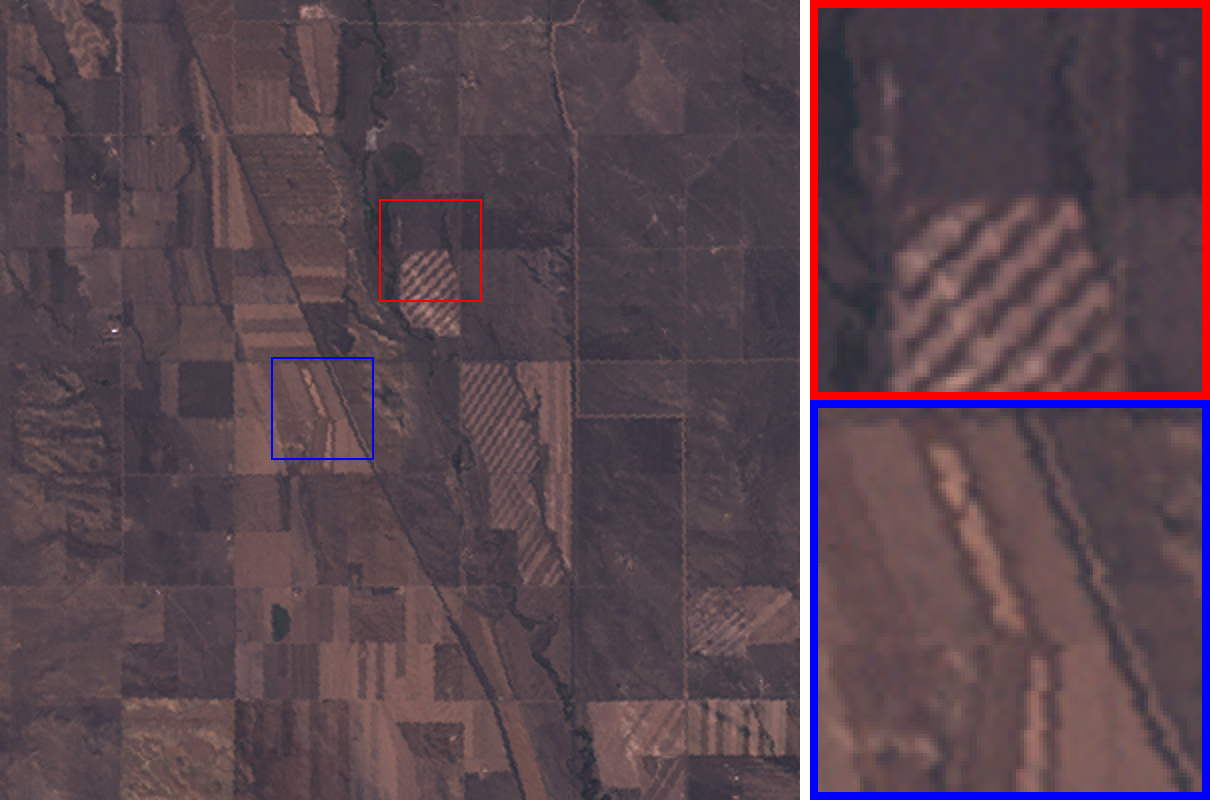}
    \caption{MS image}
    \label{fig:comparison_pansharpening_a}
\end{subfigure}
\begin{subfigure}{0.3\linewidth}
    \includegraphics[width=\linewidth]{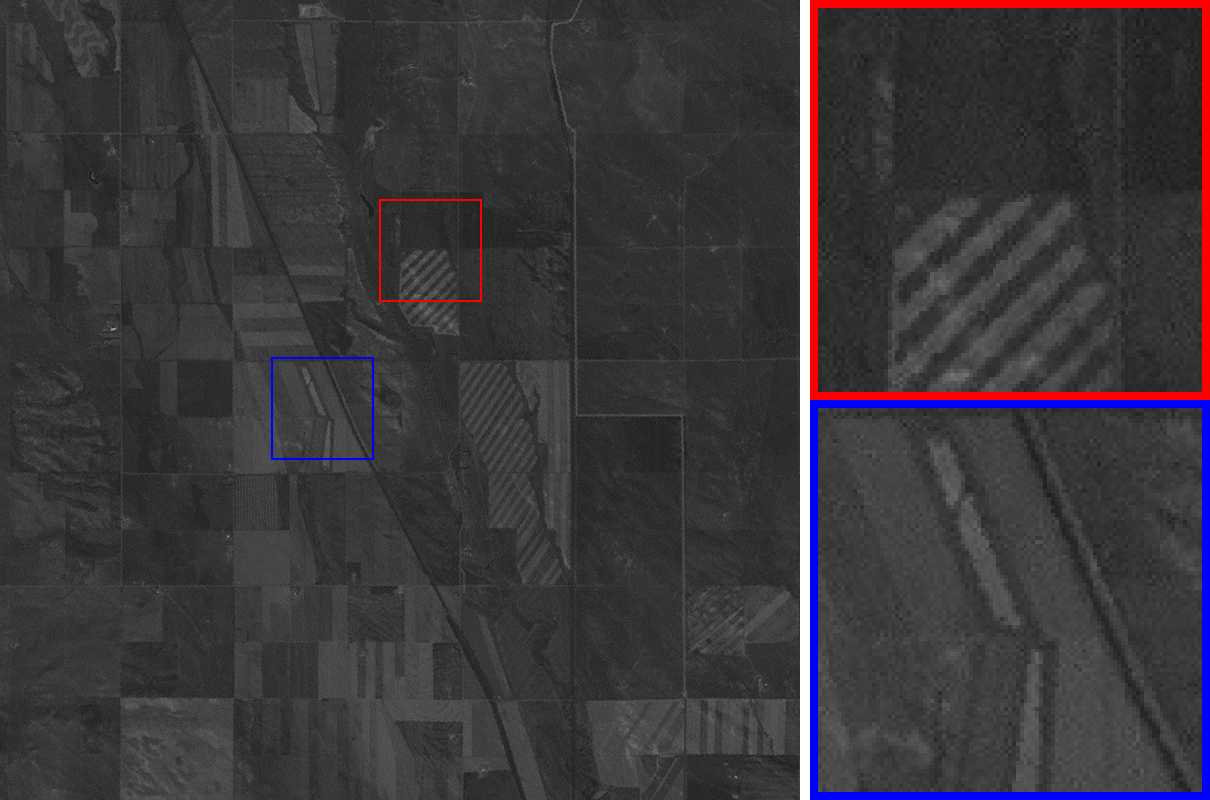}
    \caption{PAN image}
    \label{fig:comparison_pansharpening_b}
\end{subfigure}
\begin{subfigure}{0.3\linewidth}
    \includegraphics[width=\linewidth]{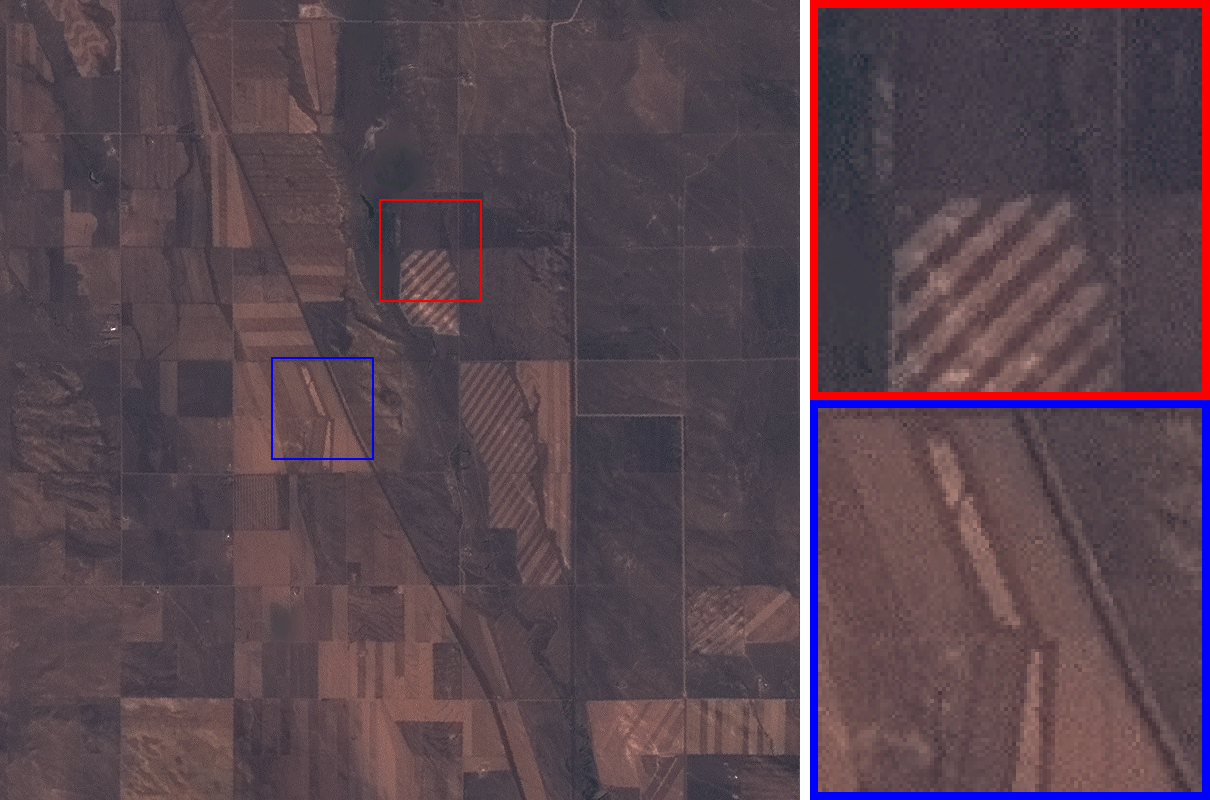}
    \caption{Proposed, ERGAS $= 6.53 $}
    \label{fig:comparison_pansharpening_c}
\end{subfigure}

\begin{subfigure}{0.3\linewidth}
    \includegraphics[width=\linewidth]{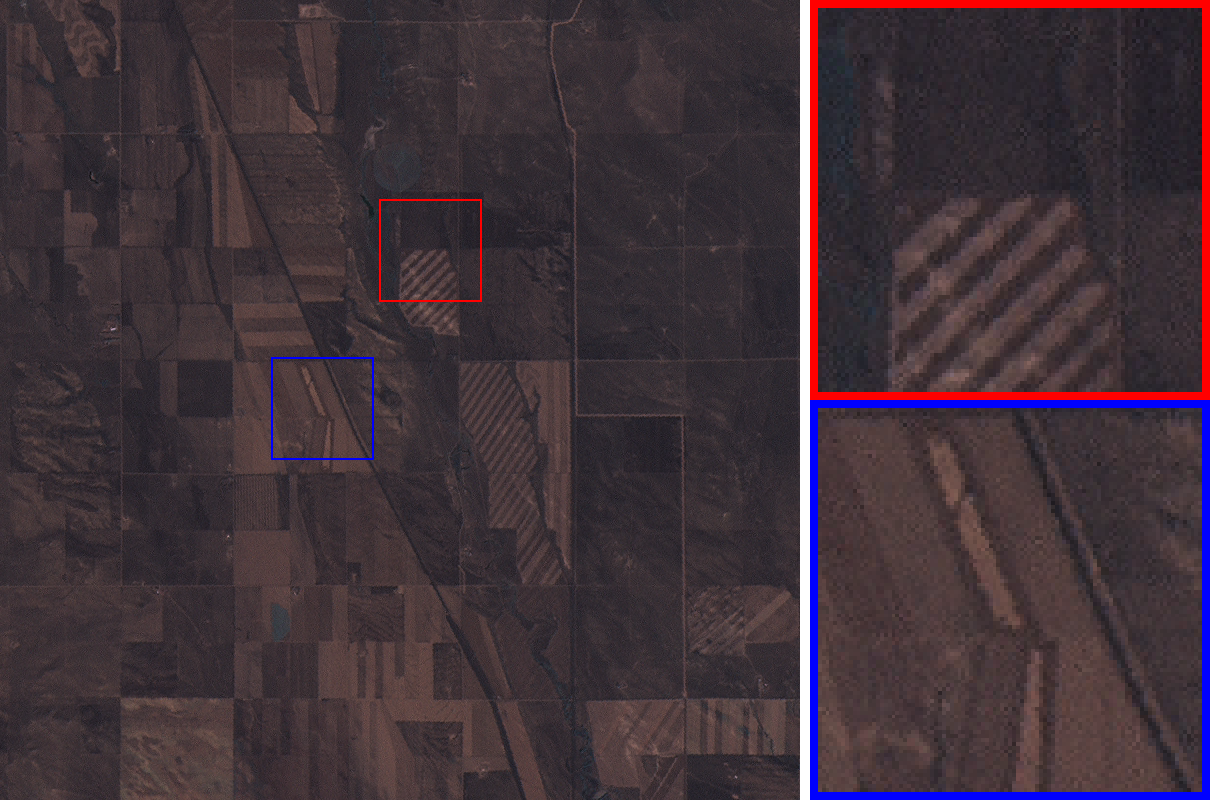}
    \caption{BT \cite{Tu2005}, ERGAS $ = 30.25 $}
    \label{fig:comparison_pansharpening_d}
\end{subfigure}
\begin{subfigure}{0.3\linewidth}
    \includegraphics[width=\linewidth]{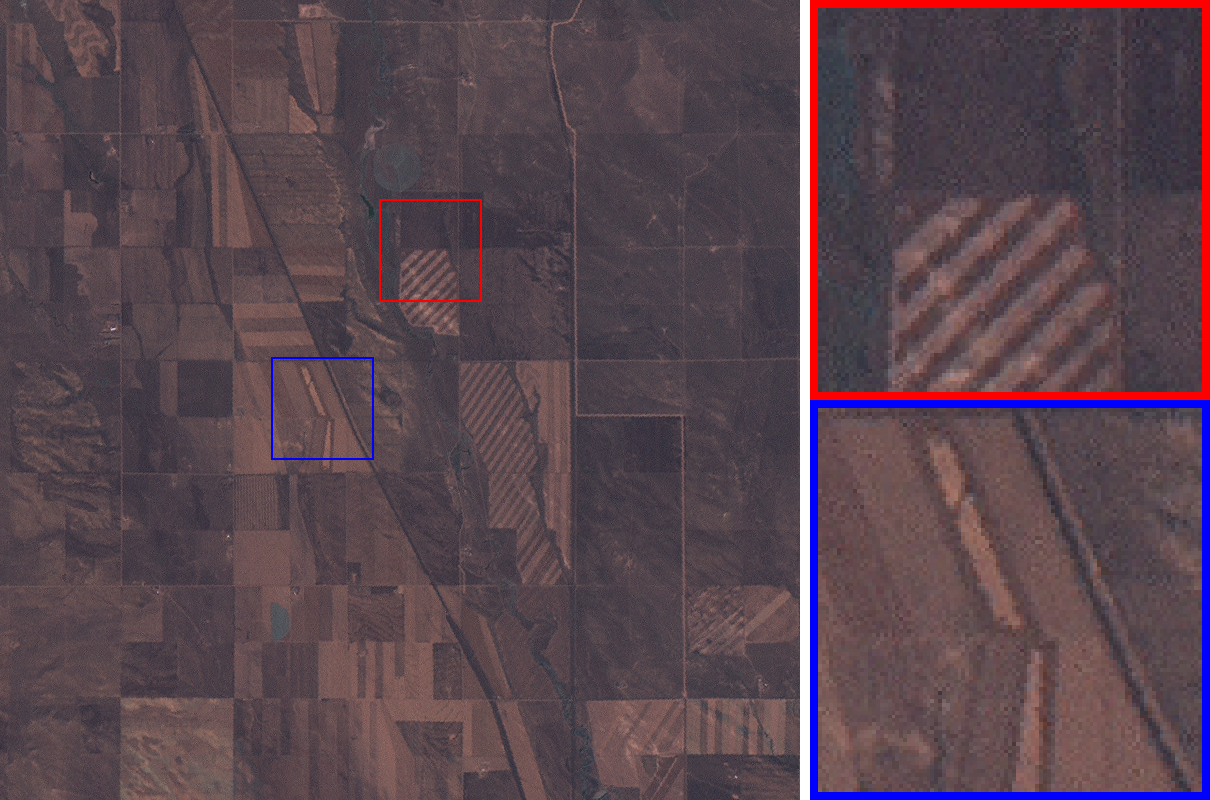}
    \caption{IHS \cite{tu2001new}, ERGAS $ = 6.77$}
\end{subfigure}
\begin{subfigure}{0.3\linewidth}
    \includegraphics[width=\linewidth]{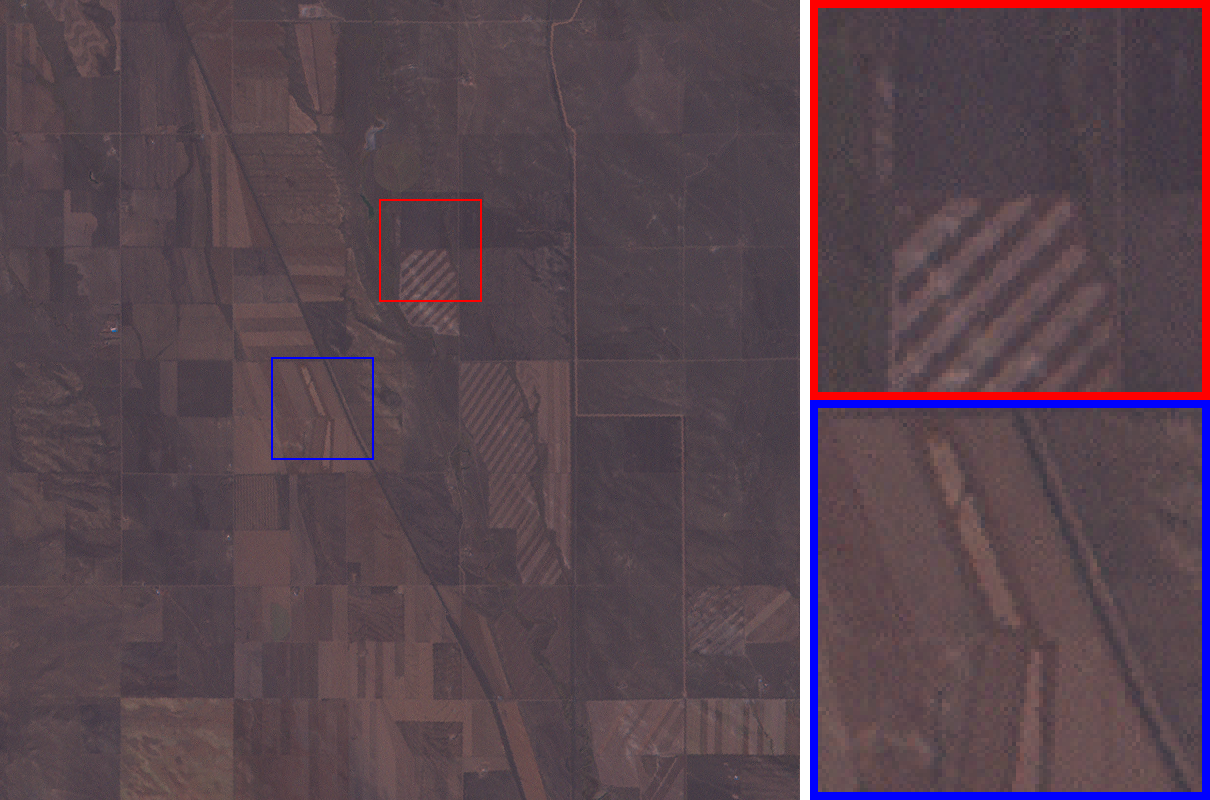}
\caption{PCA \cite{shah2008efficient}, ERGAS $ = 8.78 $}
\end{subfigure}

\begin{subfigure}{0.3\linewidth}
    \includegraphics[width=\linewidth]{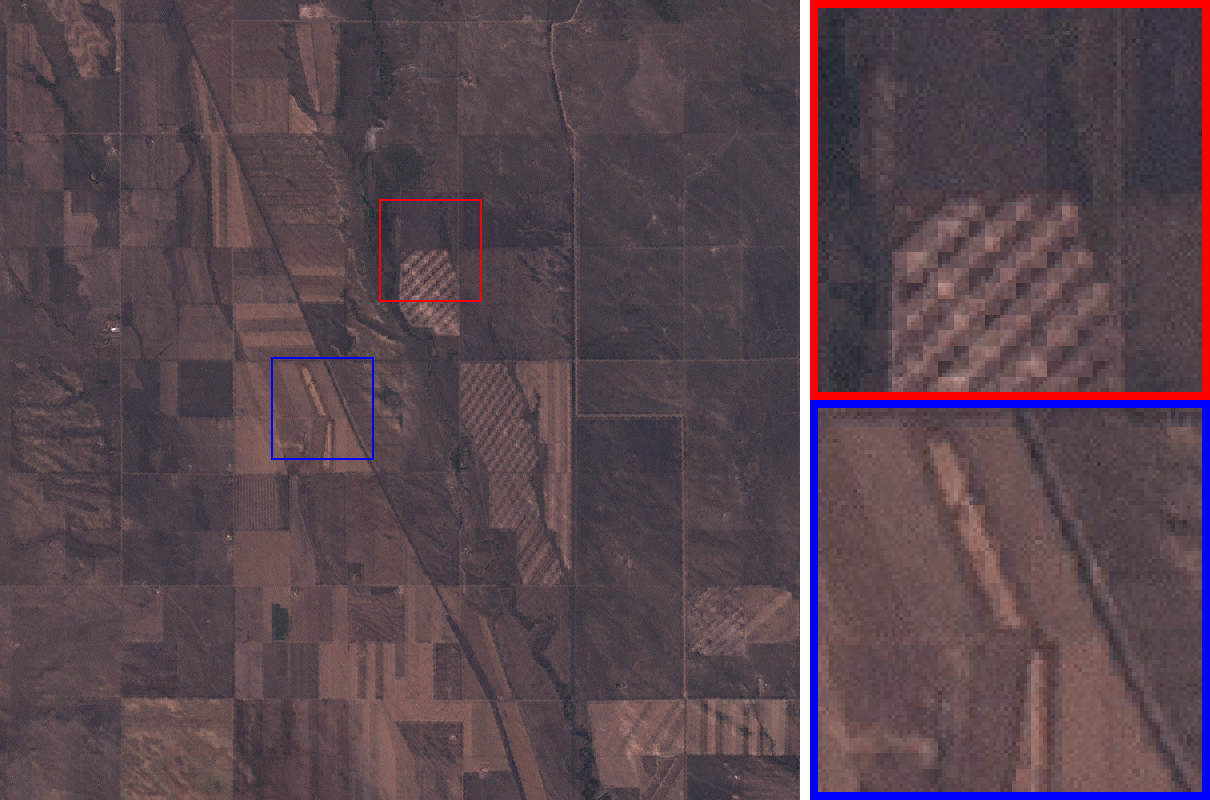}
    \caption{WT \cite{nunez1999multiresolution}, ERGAS $ = 4.64$}
\end{subfigure}
\begin{subfigure}{0.3\linewidth}
    \includegraphics[width=\linewidth]{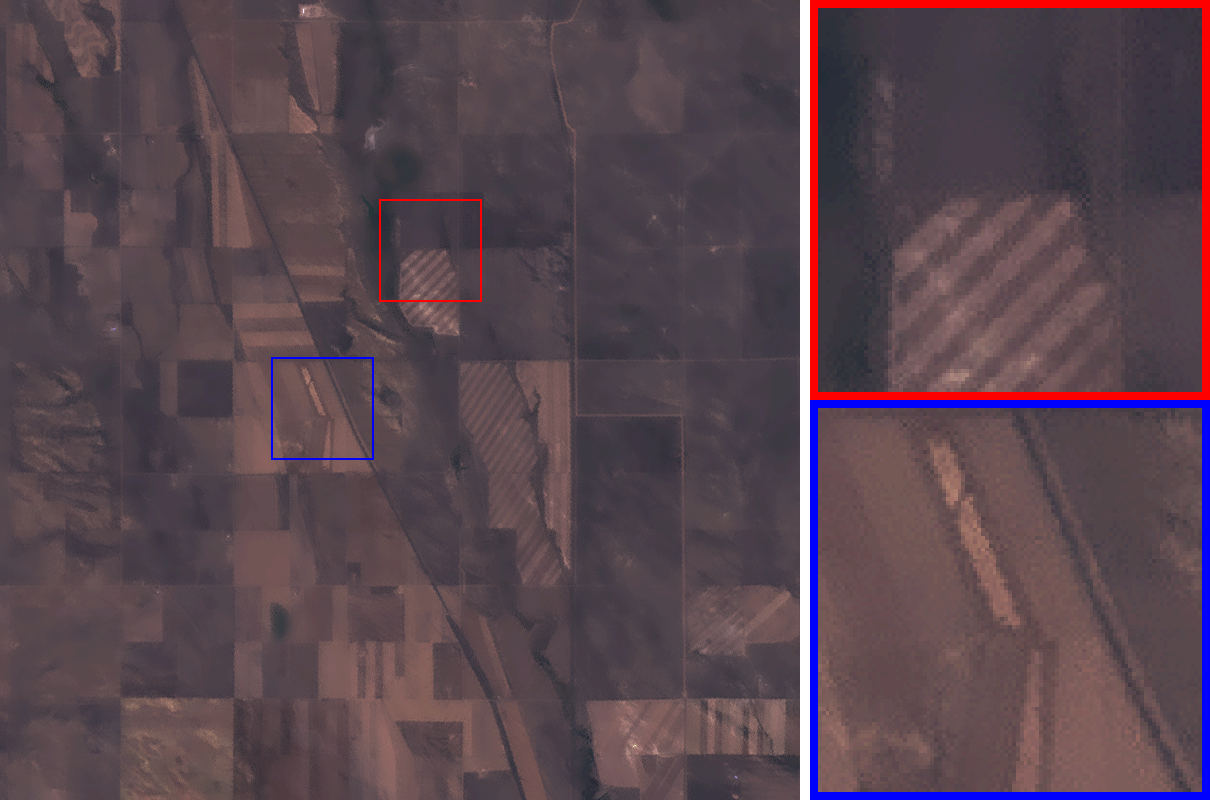}
    \caption{GF \cite{Qu2017}, ERGAS $ = 5.83 $}
\end{subfigure}
\begin{subfigure}{0.3\linewidth}
    \includegraphics[width=\linewidth]{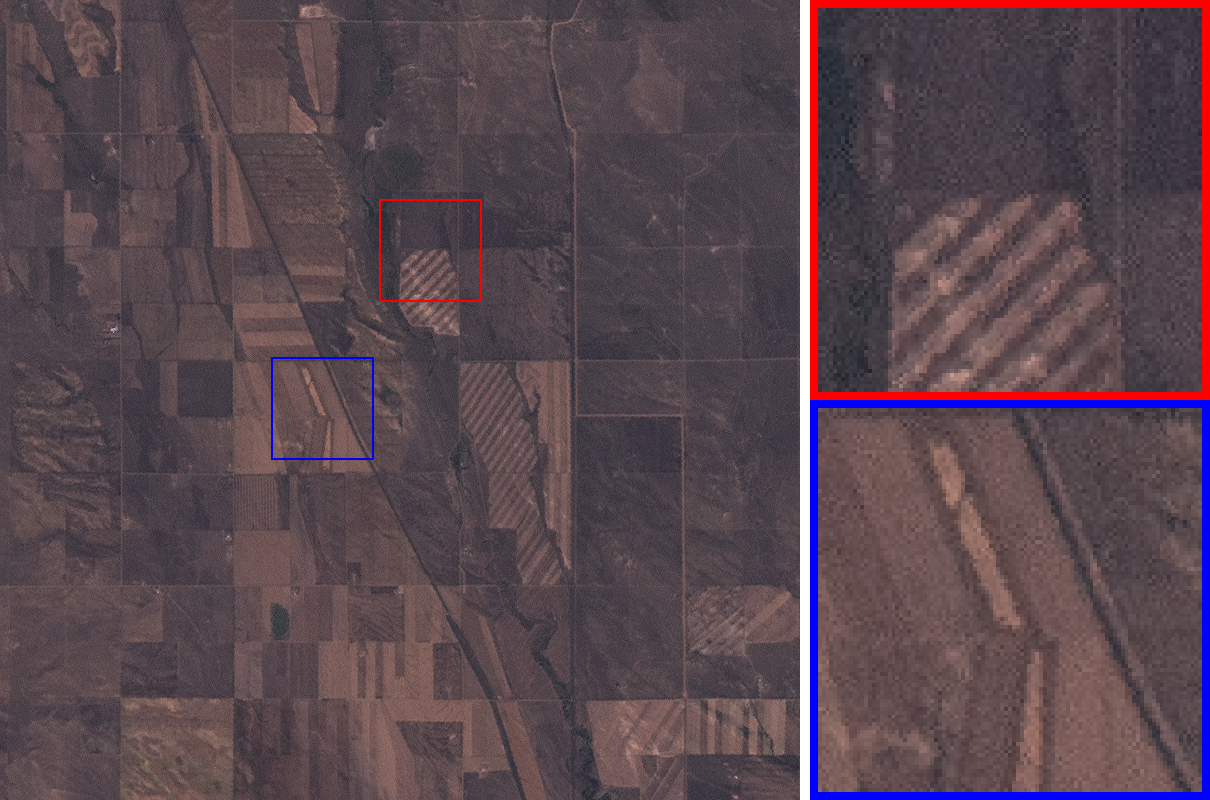}
\caption{P+XS \cite{ballester2006variational}, ERGAS $ = 5.58 $}
\end{subfigure}

\caption{Pan-sharpening results and comparison with other methods. (a) Original low resolution multi spectral image. (b) High resolution panchromatic image. (c) Proposed filter. (d) to (i) Other state of the art methods. The proposed filter successfully transfers the details from the high spatial resolution panchromatic image to the up-sampled multi spectral image while preserving the spectral resolution. }
\label{fig:comparison_pansharpening}
\end{figure*}

Pan-sharpening techniques have been continuously developed over the years. Some frequently used techniques include: Brovery transform (BT)\cite{Tu2005}, intensity-hue-saturation (IHS) \cite{tu2001new}, principal component analysis (PCA) \cite{shah2008efficient}, wavelet transform (WT) \cite{nunez1999multiresolution, otazu2005introduction}, guided filter based methods such as \cite{Liu2016, Qu2017}, and P+XS \cite{ballester2006variational}. We use some of these methods to compare with the result of the proposed algorithm. 

The proposed method consists of 3 steps similar to the GF-based approach \cite{Qu2017}. The MS image is first up-sampled using nearest neighbor interpolation. The result is then processed by the proposed filter which uses the PAN image as a guidance to transfer the high spatial resolution information from the guidance image to the low resolution MS image. The last step is a histogram matching on the result image using the original MS image as a reference. 

\figref{fig:comparison_pansharpening_a} and \figref{fig:comparison_pansharpening_b} show the up-sampled MS image and the PAN images respectively. The images are from United States Geological Survey database\footnote{\href{https://earthexplorer.usgs.gov/}{https://earthexplorer.usgs.gov/}}. \figref{fig:comparison_pansharpening_c} shows the output of the proposed algorithm using $r=11, \epsilon = 0.1, \kappa = 1.2, scale = 0.5$, the details of the PAN image are successfully transferred to the up-sampled MS image and the spectral resolution is preserved. In the second and third rows of \figref{fig:comparison_pansharpening} we present the results of the Pan-sharpening process using BT, IHS, PCA, WT, GF and P+XS algorithms. We can observe that our method produces sharp and high contrast results with large spatial and spectral information. 

We use ERGAS \cite{wald2000quality} to quantify the quality of a Pan-sharpened image. ERGAS is a metric that calculates the spectral distortion. Ideally, its value should be zero. The ERGAS values for results produced by different methods are shown in the caption of \figref{fig:comparison_pansharpening}. We can see that the result of the proposed filter is of about the same quality as those produced by state-of-the-art methods. Thus, the proposed filter is a new tool for Pan-sharpening with the ability to perform pixel-wise sharpening or smoothing.

\section{Conclusions}

Smoothing and sharpening are two fundamental operations in image processing. They are usually related through the unsharp masking algorithm. In this paper, we have developed a new filter which can perform smoothing and sharpening depending on the setting of a parameter $\kappa$. The filter is a smoothing filter or a sharpening filter when $0<\kappa<1$ or $\kappa>1$. The systematic unification of these two operations in one filter is based on (a) a new Laplacian based filter formulation which unifies the smoothing and sharpening operations, (b) a patch interpolation model similar to the  guided filter which provides the edge-awareness capability, and (c) the generalized Gamma distribution as the prior for parameter estimation. As a result the filter allows pixel-adaptive image smoothing/sharpening by adapting $\kappa$ to local characteristics such as texture, depth, and blurriness. Based on the patch interpolation model, the proposed filter uses the guidance information in two ways. In self-guidance the proposed filter uses information of the image to be processed and has the ability to use other information to adapt $\kappa$. In external-guidance, the filter is similar to the guided filter, but has an extra ability of adaptive smoothing-sharpening. In addition, the proposed filter has the desirable edge-awareness property which retains sharp edges in smoothing and does not suffer from the halo effect in sharpening.

Using the filter in self-guidance we have developed adaptive smoothing-sharpening algorithms based on information of texture, depth and blurriness to enhance human face images, to create the effect of shallow depth of field, to perform adaptive processing based on local blurriness, and to pre-process an image to achieve better seam carving results. Using the filter in external guidance, we have combined images of under flash and no-flash conditions, producing much better results than those produced by using the guided filter. We have also demonstrated the successful application of the filter to solve the Pan-sharpening problem which combines information from multi-spectral images with a panchromatic image.

\section*{Appendix}
A brute-force implementation of the proposed filter in MATLAB is presented as the following code. We assume parameters such as: patch radius ($r$), Kappa ($\kappa$), Epsilon $(\epsilon)$, and Scale $(s)$, are provided by the user.  
\vspace{0.5cm}
%
%
%
%
  
\definecolor{mblue}{rgb}{0,0,1} 
\definecolor{mgreen}{rgb}{0.13333,0.5451,0.13333} 
\definecolor{mred}{rgb}{0.62745,0.12549,0.94118} 
\definecolor{mgrey}{rgb}{0.5,0.5,0.5} 
\definecolor{mdarkgrey}{rgb}{0.25,0.25,0.25} 
  
\DefineShortVerb[fontfamily=courier,fontseries=m]{\$} 
\DefineShortVerb[fontfamily=courier,fontseries=b]{\#} 
  
\noindent                      
 $pad=$\color{mred}$'symmetric'$\color{black}$;$\\
 $N=(2*r+1)^2;$\\
 $h=ones(2*r+1)/N;$\\
 \color{mgreen}$
 $mu=imfilter(I,h,pad);$\\
 $$\color{mgreen}$
 $nu=imfilter(G,h,pad);$\\
 \color{mgreen}$
 $phi=imfilter(I.*G,h,pad)-mu.*nu;$\\
 $$\color{mgreen}$
 $vS=imfilter(G.*G,h,pad)-nu.*nu;$\\
 $a=phi./(vS+Epsilon);$\\
 $Beta=(a+sign(phi).*sqrt(a.^2+4*kappa...$\\
 $    *Epsilon./(vS+Epsilon)))/2;$\\
 \color{mgreen}$
 $w=vS./(s*mean(vS(:)));$\\
 $w=1./(1+w.^2);$\\
 $nor=imfilter(w,h,pad);$\\
 \color{mgreen}$
 $A=imfilter(Beta.*w,h,pad);$\\
 $B=imfilter((mu-Beta.*nu).*w,h,pad);$\\
 $J=(G.*A+B)./nor;$\\ 
  
\UndefineShortVerb{\$} 
\UndefineShortVerb{\#}


\end{document}